\documentclass[namedreferences]{%
SolarPhysics}  
\usepackage[optionalrh]{%
spr-sola-addons} 
\usepackage{graphicx}        
\usepackage{color}           
\usepackage{url}             

\definecolor{blue}{rgb}{0,0,1} 

\newcommand{\citep}{\cite}
\newcommand{\citet}{\inlinecite}
\newcommand{\citealt}{\opencite}

\newcommand{\be}{\begin{equation}}
\newcommand{\ee}{\end{equation}}
\newcommand{\bea}{\begin{eqnarray}}
\newcommand{\eea}{\end{eqnarray}}
\newcommand{\beax}{\begin{eqnarray*}}
\newcommand{\eeax}{\end{eqnarray*}}
\newcommand{\ba}{\begin{array}}
\newcommand{\ea}{\end{array}}
\newcommand{\blc}{\begin{list}{$\circ$}{}}
\newcommand{\blb}{\begin{list}{$\bullet$}{}}
\newcommand{\el}{\end{list}}
\newcommand{\ben}{\begin{enumerate}}
\newcommand{\een}{\end{enumerate}}

\newcommand{\etal}{{\it et al.}}

\renewcommand{\vec}[1]{ {\mathbf #1} }
\newcommand{\uvec}[1]{ \hat{\mathbf #1} }

\newcommand{\avec}{ \vec A}

\newcommand{\evec}{ \vec E}
\newcommand{\Uvec}{ \vec u}
\newcommand{\bvec}{ \vec B}
\newcommand{\vvec}{ \vec v}

\newcommand{\aap}{    {\it Astron. Astrophys.}}

\newcommand{\apj}{    {\it Astrophys. J.}}
\newcommand{\apjl}{   {\it Astrophys. J.}}

\newcommand{\jgr}{    {\it J. Geophys. Res.}}

\newcommand{\solphys}{{\it Solar Phys.}}
\newcommand{\ssr}{    {\it Space Sci. Rev.}}

\begin{document}

\begin{article}

\begin{opening}

\title{Photospheric Magnetic Evolution in the WHI Active Regions}

\author{B.T.~\surname{Welsch}$^{1}$\sep
  S.~\surname{Christe}$^{2}$\sep
  J.M.~\surname{McTiernan}$^{1}$}

\runningauthor{B.T. Welsch, J.M. McTiernan, and S.D. Christe}
\runningtitle{WHI Photospheric Magnetic Activity}

\institute{$^{1}$ Space Sciences Laboratory, University of California, 
  7 Gauss Way, Berkeley, CA 94720-7450, U.S.A. \\
  email: \url{welsch@ssl.berkeley.edu}}

\institute{$^{2}$NASA/Goddard Space Flight Center, 
Greenbelt, MD 20771, U.S.A.}

\begin{abstract}
Sequences of line-of-sight (LOS) magnetograms recorded by the
{\it Michelson-Doppler Imager} are used to quantitatively characterize
photospheric magnetic structure and evolution in three active regions
that rotated across the Sun's disk during the Whole Heliosphere
Interval (WHI), in an attempt to relate the photospheric magnetic
properties of these active regions to flares and coronal mass
ejections (CMEs).
Several approaches are used in our analysis, on scales ranging from
whole active regions, to magnetic features, to supergranular scales,
and, finally, to individual pixels.  
We calculated several parameterizations of magnetic structure and
evolution that have previously been associated with flare and CME
activity, including total unsigned magnetic flux, magnetic flux near
polarity inversion lines, amount of cancelled flux, the ``proxy
Poynting flux,'' and helicity flux.
To catalog flare events, we used flare lists derived from both GOES
and RHESSI observations.
By most such measures, AR 10988 should have been the most flare- and
CME-productive active region, and AR 10989 the least.  Observations,
however, were not consistent with this expectation: ARs 10988 and
10989 produced similar numbers of flares, and AR 10989 also
produced a few CMEs.
These results highlight present limitations of statistics-based flare
and CME forecasting tools that rely upon line-of-sight photospheric
magnetic data alone.
\end{abstract}
\end{opening}


\section{Characterizing Photospheric Magnetic Evolution}

Evolution of magnetic fields in solar active regions (ARs) affects the
heliosphere in many ways.  In the low corona, sudden magnetic
evolution in flares and coronal mass ejections (CMEs) --- which arise
on timescales of minutes to hours --- can launch powerful disturbances
into interplanetary space.  More gradual magnetic evolution, such as
the effective diffusion and transport of active region magnetic flux over
the photosphere, also drives evolution in heliospheric structure
(affecting, e.g., the structure of the streamer belt and the
heliospheric current sheet), on typical timescales of weeks and
months.  Consequently, studying the evolution of active-region
magnetic fields is crucial to understanding the structure and dynamics
of the heliosphere, beyond understanding solar activity itself.

Quantifying evolution of magnetic fields in photospheric magnetograms,
in particular, can provide insights into heliospheric evolution.
Photospheric magnetograms reveal cross sections of the large-scale
structure of active regions, which extend from the solar interior out
into the corona.
The coupling of the photospheric field to the coronal field implies
that magnetic evolution at the photosphere will induce evolution in
the coronal field.  Note, however, that the coronal magnetic field can
evolve independently of the photospheric field; for instance, MHD
instabilities that arise in the coronal field might trigger flares or
CMEs ({\it e.g.} \citealt{Forbes2000}).


Here, we present analyses of several aspects of photospheric magnetic
evolution in the three active regions (NOAA ARs 10987, 10988, and
10989) that crossed the solar disk during the Whole Heliosphere
Interval (WHI), as observed in line-of-sight (LOS) magnetograms
recorded with the {\it Michelson-Doppler Imager} (MDI) instrument
\citep{Scherrer1995} on the SOHO satellite.  The data are
described in Section \ref{sec:data}, our analysis methods and results are
discussed in Section \ref{sec:methods}, and we conclude by discussing the
implications of these results in Section \ref{sec:discussion}.

\section{Data}
\label{sec:data}

\subsection{Magnetograms}
\label{subsec:mgrams}

\begin{figure}
  \centerline{\includegraphics[width=0.875\textwidth]{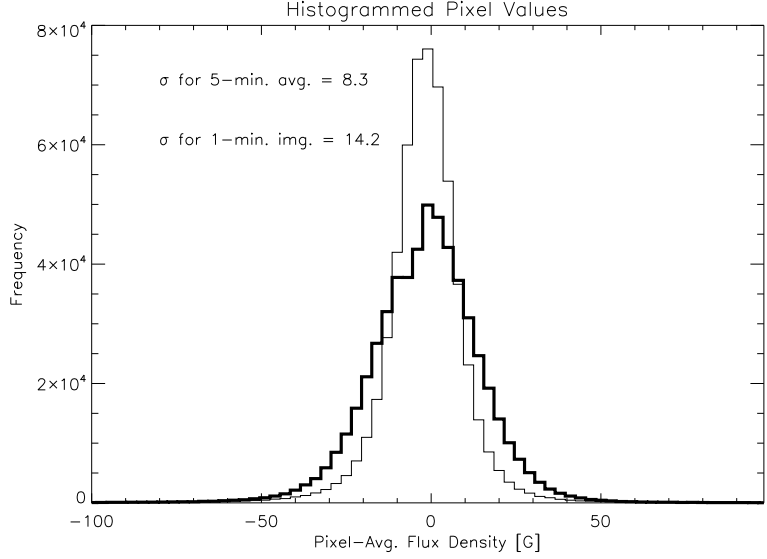}}
  \caption{Histograms of pixel-average field strengths in
    single- (thick) and five-measurement (thin)
    magnetograms. Gaussian fits to the histograms can be
    used to estimate the noise levels in each, which are
    $\sim 14$G and $\sim 8$G, respectively.}
  \label{fig:hist}
\end{figure}

Our analysis of photospheric magnetic evolution in the WHI ARs began
with selection of magnetograms from the database of full-disk,
line-of-sight, 96-minute-cadence, Level 1.8.2 magnetograms from MDI
(\url{http://soi.stanford.edu/magnetic/Lev1.8/}).  These magnetograms
are formed either from a single measurement, nominally of 30 seconds,
or by summing measurements over five minutes.  The 30-second
magnetograms are much more common in the dataset than the five-minute
magnetograms; and the former are noisier.  To estimate the noise
levels in each, we fit Gaussians to histograms of field strengths
(which are actually pixel-averaged flux densities) in one sample
magnetogram of each type, and took the noise level as 1$\sigma_{\rm
  fit}$.  Each sample magnetogram was smoothed with a 2D, three-pixel
boxcar prior to histogramming into 3 G bins.  While we expect
noise to dominate the weakest-field pixels, and would therefore prefer
to fit only the ``core'' of each histogram \citep{Hagenaar1999}, in
practice using ranges smaller than $\pm30$G resulted in poor fits.
Results fitting this range of field strengths differed little from
those derived by fitting the whole histogram.  Whole-histogram
$\sigma_{\rm fit}$ values were $\simeq 14$G and $\simeq 8$G for single
and five-measurement magnetograms, respectively; the sample histograms
used are plotted in Figure \ref{fig:hist}.  Given the scarcity of the
five-minute magnetograms, we assume a single noise level of $14$G for
simplicity.

The MDI instrument measures the LOS field strength $[B_{\rm LOS}]$
averaged over each pixel. Tracking the evolution of the radial
magnetic field $[B_R]$ requires estimating $B_R$, since only $B_{\rm
  LOS}$ was observed.  We therefore assumed the magnetic field was
radial, and applied cosine corrections to the LOS field in each pixel,
$B_R = B_{\rm LOS}/\cos(\gamma)$, where $\gamma$ is the heliocentric
angle from disk center to each pixel.  To compensate for
foreshortening, triangulation was used to interpolate the $B_R$ data
--- regularly gridded in the plane of the sky, but irregularly gridded
in spherical coordinates $(\theta,\phi)$ on the solar surface --- onto
points $(x,y)$ corresponding to a regularly gridded Mercator
projection of the spherical surface, following \citet{Welsch2009}.
This projection is conformal (and so locally preserves shape), which
is necessary to ensure displacements measured by our tracking methods
were not biased in direction. This reprojection distorts length
scales, such that apparent displacements are exaggerated by a factor
of the secant of the apparent latitude, which we corrected after tracking.

\begin{figure}
  \centerline{\includegraphics[width=0.95\textwidth]{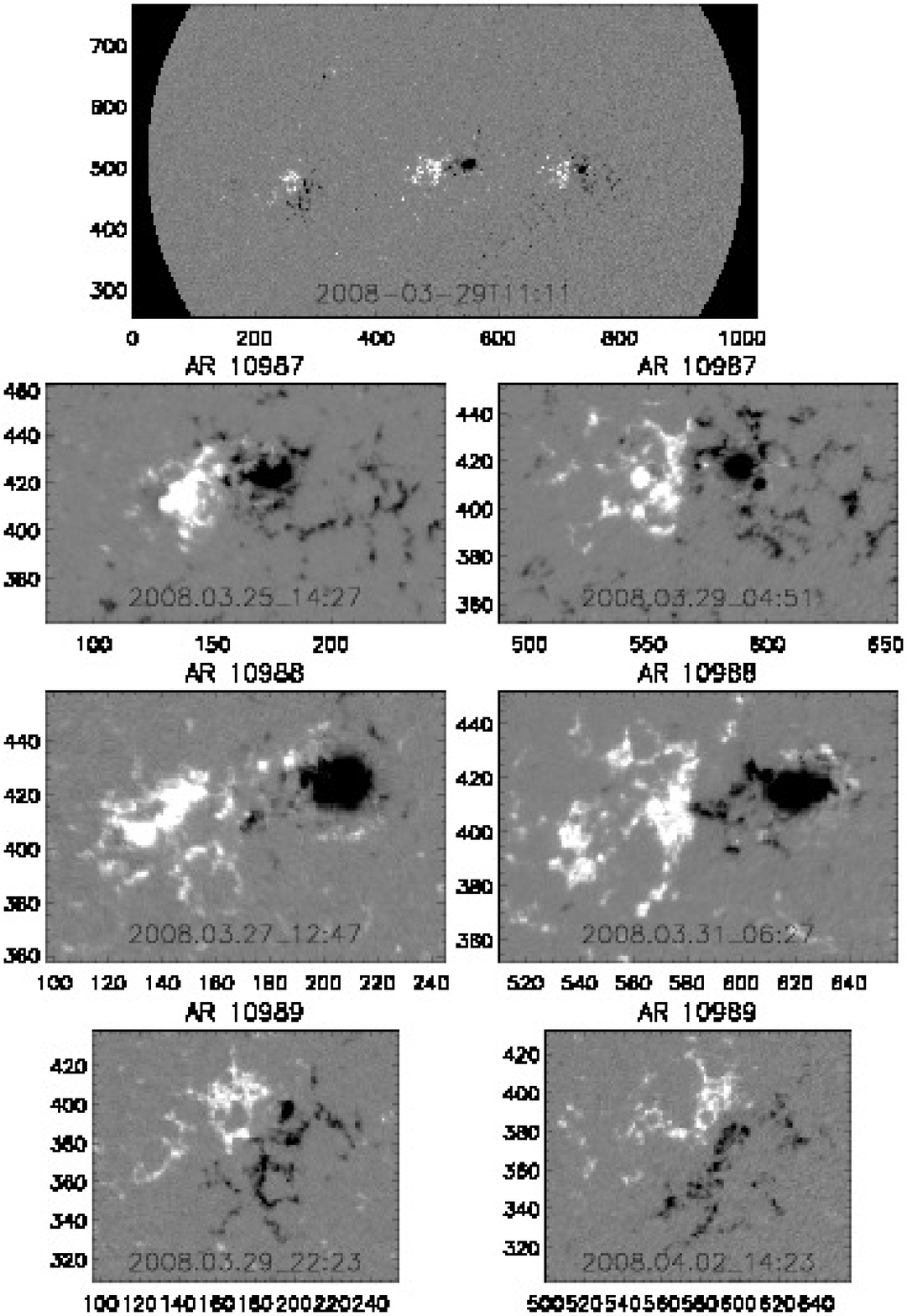}}
  \caption{Top: This full-disk magnetogram, cropped
      at $\pm 30^\circ$, shows all three WHI ARs (from right to left,
      10987, 10988, 10989).  The cropped magnetograms in each of the
      subsequent rows show each AR roughly one day after we began
      tracking it (left) and one day before we stopped tracking it
      (right).  Axes' labels are pixels from MDI full-disk
      magnetograms.}
  \label{fig:mgrams}
\end{figure}

Since the MDI magnetograph only measures the LOS component of the
photospheric magnetic field, we chose to analyze only magnetograms in
which the target AR was within about 45$^\circ$ of disk center.  The
first AR visible on the disk during WHI was AR 10987 on 24 March 2008,
and we started our tracking analyses with the magnetogram recorded at
14:23 UT.  Shortly thereafter, AR 10988 rotated onto the disk,
followed by AR 10989.  By 2 April, AR 10989 had no sunspots, but we
analyzed magnetograms through 4 April, when AR 10989 was nearly
45$^\circ$ from disk center, ending at 14:23 UT.  The time ranges over
which each AR was tracked are listed in Table \ref{tab:tracktimes}.
The top-most magnetogram in Figure \ref{fig:mgrams} shows all three
WHI ARs on the disk.  The magnetograms in each of the subsequent rows
of the same Figure show each WHI AR roughly one day after we began
tracking it (left) and one day before we stopped tracking it (right).

\begin{table}
\caption{Intervals over which photospheric magnetic fields of WHI ARs
  were tracked on the central disk.}
\label{tab:tracktimes}
\begin{tabular}{cccc}
  \hline                   
NOAA AR & Tracking Start & Tracking End  \\
  \hline
10987 & 2008 Mar 24, 14:23UT & 2008 Mar 30, 04:47UT \\ 
10988 & 2008 Mar 26, 12:48UT & 2008 Apr 01, 06:23UT \\
10989 & 2008 Mar 28, 22:23UT & 2008 Apr 03, 14:23UT \\
  \hline
\end{tabular}
\end{table}

\subsection{Flare and CME Activity}
\label{subsec:flares}

Our primary goal is to relate properties of magnetic structure and
evolution at the photosphere to energy release in the corona, in the
form of flares and CMEs.  

\citet{Webb2011} used several sources of data to compile a
comprehensive list of CMEs over the WHI, and to determine their source
regions.  We used their results to determine CME productivity of the
WHI ARs during the time intervals shown in Table \ref{tab:flareobs}.
Only AR 10989 was CME-productive while tracked; it produced two during
its tracking interval (and a further two in the hours just outside the
interval over which it was tracked).

While CMEs are generally responsible for the strongest heliospheric
disturbances \citep{Gosling1993}, flares can also affect the
heliosphere.  As detailed below, each WHI AR only produced relatively
small flares during the interval over which it was tracked.  But even
weak flares can affect the heliosphere: for instance, small hard X-ray
(HXR) bursts are often associated with interplanetary Type III radio
bursts (e.g. \citealt{Christe2008b}), which arise from electrons escaping
into the heliosphere (where they have been observed {\em in situ}).

To characterize flare activity in the WHI ARs, we first consulted
records of soft X-ray (SXR) flare emission observed with the GOES
satellites in the flare catalog maintained by NOAA (\url{
  ftp://ftp.ngdc.noaa.gov/STP/SOLAR\_DATA/SOLAR_FLARES/FLARES_XRAY/}),
and found that most of the relatively weak flares during WHI were not
attributed to any source region, meaning this catalog did not
accurately characterize each AR's flare activity.  Next, we consulted
Sam Freeland's Latest Events Archive (LEA:
\url{http://www.lmsal.com/solarsoft/latest_events_archive.html})
which differences EUV images to determine flare-source locations.  The
LEA flare list identified the source ARs for many events that lacked
source ARs in the GOES catalog.  In the three cases where the NOAA and
LEA source attributions disagreed, manual inspection of LEA difference
images suggested the LEA attributions were probably correct.  Some
flares in the LEA list were incorrectly attributed to either: {\it i)}
remnant ARs from previous rotations that were due to rotate back to
the flares' positions when they occurred, or {\it ii)} ARs farther
West than the true source AR.  Both errors might arise from using
active region locations from Solar Region Summaries (SRSs), jointly
prepared once per day by NOAA's Space Weather Prediction Center (SWPC;
\url{http://www.swpc.noaa.gov/ftpdir/forecasts/SRS/README}) and the
U.S. Air Force, that were issued many hours before flare time.  While
source latitudes and longitudes in the LEA list appear accurate for
the events we studied, AR source attributions from both the NOAA and
LEA event lists contain errors.

A more robust approach to automatic identification of flares' source
active regions uses data from the RHESSI satellite \citep{Lin2002},
which is capable of localizing the source regions of hard X-rays
(HXRs) emitted in flares (as well as imaging HXR emission in
high-fluence flares).  
%
%
Recent improvements in the algorithms used to identify microflares
(short bursts of HXR emission in RHESSI's 6\,--\,12 keV energy band;
see, e.g., \citealt{Christe2008}), which are invariably associated with
active regions, has enabled their inclusion in the most recent RHESSI
flare catalog
(\url{http://hessi.ssl.berkeley.edu/hessidata/dbase/hessi_flare_list.txt}).
To find the source AR for each flare in the RHESSI catalog, disk
positions for the WHI ARs from the daily SRSs were used to develop
linear models of for each AR's apparent latitude and longitude as a
function of time.  This approach enabled extrapolating the positions
of ARs both before they were visible near the east limb and recognized
as new ARs, and after they passed beyond the west limb.  Flares were
automatically associated by proximity to the nearest predicted AR
location.  Visual inspection of each automatic association confirmed
that this method correctly assigned above-the-limb flares to ARs
which, though not visible, were at or near the limb.  
%
%
%
Lightcurves from GOES were then searched to identify the transient
X-ray enhancement associated with each RHESSI flare, and the
enhancement's peak intensity was recorded. 

\begin{figure}
  \centerline{\includegraphics{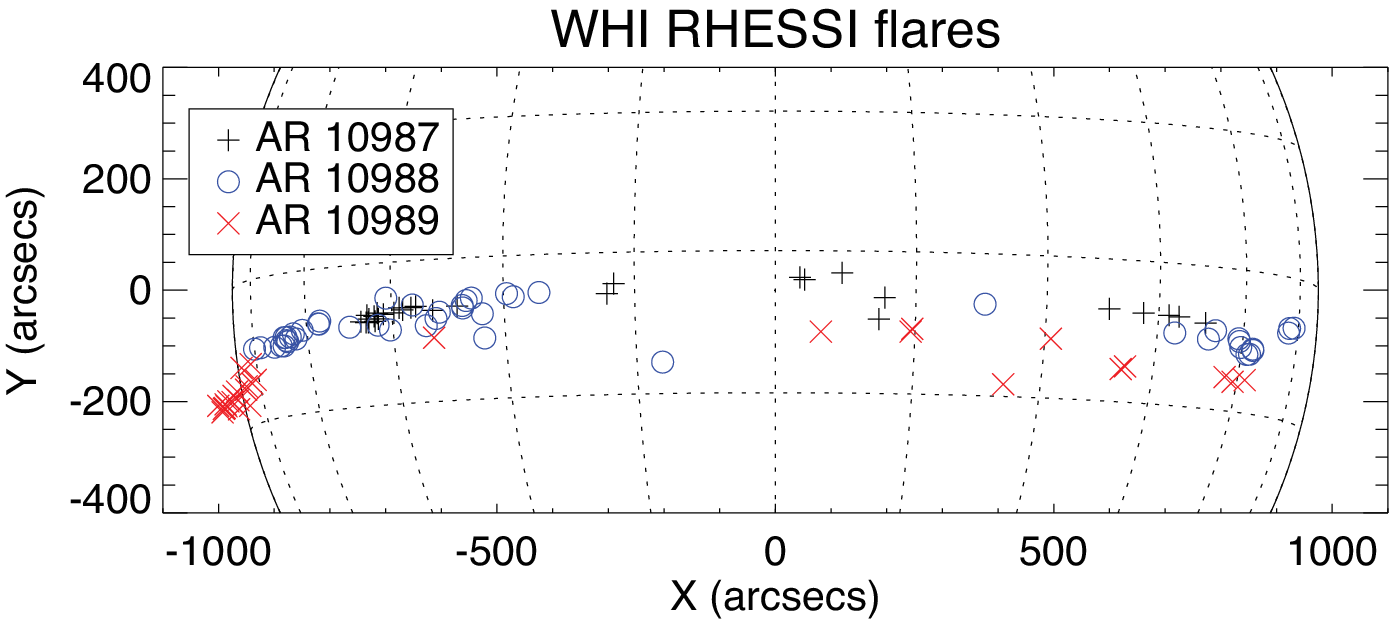}}
  \caption{Positions and source ARs for flares identified by RHESSI.}
  \label{fig:globe}
\end{figure}

Flares and source AR attributions are listed in Table
\ref{tab:christe}, and plotted in Figure \ref{fig:globe}, which shows
the position of each RHESSI flare with respect to the solar disk, and
its AR association.
We note that most of the flares for AR 10988 occurred on the eastern
disk, and many of the flares for 10989 occured above the east limb.
Consequently, the LOS magnetograms from MDI cannot tell us much about
the photospheric magnetic structure of these ARs during their most
flare-productive intervals.

\begin{table}
\caption{RHESSI HXR bursts in the 6–12 keV channel during the WHI
  interval from the RHESSI flare list. Only those events with valid
  positions are shown in this list. Positions are listed in
  heliographic coordinates (in degrees) if on the disk or in
  heliocentric coordinates (in arcseconds) if above the limb. The
  associated GOES class is also shown for each event; some events
  occurred during GOES data gaps.}
\label{tab:christe}
\begin{tabular}{rlcccccl}
  \hline                   
\# & Date & Start & Stop & Peak & Coordinates & GOES Class & Source AR \\
  \hline
   1 & 2008 Mar 23 & 18:51 & 18:51 & 18:51 & S08E79 & A7.9 &  10988 \\ 
   2 & 2008 Mar 23 & 23:38 & 23:41 & 23:39 & S08E50 & A6.5 &  10987 \\ 
   3 & 2008 Mar 23 & 23:44 & 23:48 & 23:45 & S08E51 & B1.4 &  10987 \\ 
   4 & 2008 Mar 23 & 23:51 & 23:51 & 23:51 & S07E50 & A6.4 &  10987 \\ 
   5 & 2008 Mar 24 & 00:24 & 00:25 & 00:25 & S08E50 & B1.0 &  10987 \\ 
   6 & 2008 Mar 24 & 01:14 & 01:16 & 01:15 & S07E48 & B1.0 &  10987 \\ 
   7 & 2008 Mar 24 & 01:18 & 01:22 & 01:19 & S07E49 & B1.8 &  10987 \\ 
   8 & 2008 Mar 24 & 01:26 & 01:29 & 01:27 & S08E50 & B1.3 &  10987 \\ 
   9 & 2008 Mar 24 & 01:32 & 01:33 & 01:32 & S08E50 & B1.1 &  10987 \\ 
  10 & 2008 Mar 24 & 01:47 & 01:50 & 01:48 & S08E76 & A9.3 &  10988 \\ 
  11 & 2008 Mar 24 & 02:46 & 02:53 & 02:47 & S08E49 & B4.6 &  10987 \\ 
  12 & 2008 Mar 24 & 02:53 & 02:59 & 02:54 & S08E48 & B2.8 &  10987 \\ 
  13 & 2008 Mar 24 & 02:59 & 03:11 & 03:03 & S07E47 & B4.4 &  10987 \\ 
  14 & 2008 Mar 24 & 06:45 & 06:49 & 06:48 & S07E44 & A8.9 &  10987 \\ 
  15 & 2008 Mar 24 & 07:59 & 08:10 & 08:04 & S07E45 & B2.3 &  10987 \\ 
  16 & 2008 Mar 24 & 08:10 & 08:36 & 08:16 & S07E46 & B3.3 &  10987 \\ 
  17 & 2008 Mar 24 & 10:08 & 10:10 & 10:09 & S07E43 & A7.0 &  10987 \\ 
  18 & 2008 Mar 24 & 11:36 & 11:39 & 11:37 & S08E71 & A7.7 &  10988 \\ 
  19 & 2008 Mar 24 & 11:39 & 11:44 & 11:39 & S08E68 & A6.4 &  10988 \\ 
  20 & 2008 Mar 24 & 12:56 & 12:57 & 12:56 & S07E42 & A7.1 &  10987 \\ 
  21 & 2008 Mar 24 & 13:05 & 13:06 & 13:06 & S08E67 & A8.5 &  10988 \\ 
  22 & 2008 Mar 24 & 14:04 & 14:07 & 14:06 & S08E67 & B1.1 &  10988 \\ 
  23 & 2008 Mar 24 & 14:47 & 14:49 & 14:48 & S09E69 & A7.2 &  10988 \\ 
  24 & 2008 Mar 24 & 14:53 & 14:58 & 14:55 & S09E68 & A7.9 &  10988 \\ 
  25 & 2008 Mar 24 & 15:40 & 15:41 & 15:40 & S08E66 & A7.2 &  10988 \\ 
  26 & 2008 Mar 24 & 15:56 & 15:58 & 15:57 & S08E67 & B1.0 &  10988 \\ 
  27 & 2008 Mar 24 & 16:14 & 16:21 & 16:16 & S07E40 & B1.2 &  10987 \\ 
  28 & 2008 Mar 24 & 17:13 & 17:14 & 17:13 & S08E65 & A9.5 &  10988 \\ 
  29 & 2008 Mar 24 & 17:34 & 17:38 & 17:34 & S08E65 & A9.4 &  10988 \\ 
  30 & 2008 Mar 24 & 19:08 & 19:13 & 19:09 & -964,-197 & B1.0 &  10989 \\ 
  31 & 2008 Mar 24 & 20:52 & 20:54 & 20:53 & S07E37 & B1.3 &  10987 \\ 
  32 & 2008 Mar 24 & 21:58 & 22:04 & 21:59 & S07E63 & B1.7 &  10988 \\ 
  33 & 2008 Mar 25 & 01:59 & 02:02 & 02:00 & S07E59 & B1.4 &  10988 \\ 
  34 & 2008 Mar 25 & 04:27 & 04:30 & 04:28 & S07E59 & no data &  10988 \\ 
  35 & 2008 Mar 25 & 04:30 & 04:31 & 04:31 & S07E59 & no data &  10988 \\ 
  36 & 2008 Mar 25 & 04:50 & 05:07 & 04:54 & -966,-182 & B5.2 &  10989 \\ 
  37 & 2008 Mar 25 & 07:53 & 08:04 & 08:01 & -943,-207 & B1.8 &  10989 \\ 
  38 & 2008 Mar 25 & 09:31 & 09:35 & 09:34 & -949,-172 & B1.1 &  10989 \\ 
  39 & 2008 Mar 25 & 11:18 & 11:28 & 11:20 & -958,-139 & B1.7 &  10989 \\ 
  40 & 2008 Mar 25 & 13:18 & 13:22 & 13:21 & S08E53 & B1.2 &  10988 \\ 
  41 & 2008 Mar 25 & 14:44 & 14:47 & 14:45 & S06E47 & B1.4 &  10988 \\ 
  42 & 2008 Mar 25 & 14:49 & 14:54 & 14:52 & S11E84 & B2.3 &  10989 \\ 
  43 & 2008 Mar 25 & 16:10 & 16:12 & 16:10 & S09E82 & B1.1 &  10989 \\ 
  44 & 2008 Mar 25 & 16:12 & 16:19 & 16:15 & S08E48 & B1.1 &  10988 \\ 
  45 & 2008 Mar 25 & 17:36 & 17:41 & 17:38 & S11E81 & B2.4 &  10989 \\ 
  46 & 2008 Mar 25 & 18:44 & 19:28 & 18:51 & -950,-178 & M1.7 &  10989 \\ 
  47 & 2008 Mar 25 & 19:28 & 19:30 & 19:30 & -973,-189 & C3.4 &  10989 \\ 
\end{tabular}
\end{table}

\setcounter{table}{1} 

\begin{table}
\caption{Continued from previous page; see above for more information.}
\label{tab:christe_cont}
\begin{tabular}{rlcccccl}
  \hline                   
\# & Date & Start & Stop & Peak & Coordinates & GOES Class & Source AR \\
  \hline
  48 & 2008 Mar 25 & 20:20 & 20:28 & 20:21 & -974,-200 & B9.1 &  10989 \\ 
  49 & 2008 Mar 25 & 21:06 & 21:08 & 21:08 & -982,-199 & B4.9 &  10989 \\ 
  50 & 2008 Mar 25 & 21:56 & 22:12 & 21:57 & -987,-204 & B3.4 &  10989 \\ 
  51 & 2008 Mar 25 & 22:12 & 22:27 & 22:18 & -989,-211 & B3.1 &  10989 \\ 
  52 & 2008 Mar 25 & 22:27 & 22:43 & 22:29 & -990,-213 & B2.8 &  10989 \\ 
  53 & 2008 Mar 25 & 22:43 & 22:47 & 22:47 & -985,-208 & B2.5 &  10989 \\ 
  54 & 2008 Mar 25 & 23:32 & 23:42 & 23:35 & -992,-219 & B2.7 &  10989 \\ 
  55 & 2008 Mar 26 & 01:08 & 01:14 & 01:08 & -100,-207 & B1.5 &  10989 \\ 
  56 & 2008 Mar 26 & 01:21 & 01:23 & 01:22 & -963,-213 & B1.4 &  10989 \\ 
  57 & 2008 Mar 26 & 03:00 & 03:06 & 03:01 & S09E47 & B1.3 &  10988 \\ 
  58 & 2008 Mar 26 & 03:28 & 03:29 & 03:28 & S07E18 & B1.3 &  10987 \\ 
  59 & 2008 Mar 26 & 05:19 & 05:20 & 05:20 & S06E18 & B1.0 &  10987 \\ 
  60 & 2008 Mar 26 & 06:28 & 06:29 & 06:29 & S07E43 & B1.1 &  10988 \\ 
  61 & 2008 Mar 26 & 07:59 & 08:01 & 07:59 & S09E41 & A9.0 &  10988 \\ 
  62 & 2008 Mar 26 & 11:23 & 11:29 & 11:25 & S08E40 & B1.9 &  10988 \\ 
  63 & 2008 Mar 26 & 14:21 & 14:34 & 14:21 & S08E39 & B1.0 &  10988 \\ 
  64 & 2008 Mar 26 & 17:07 & 17:17 & 17:10 & S07E36 & B1.2 &  10988 \\ 
  65 & 2008 Mar 26 & 18:53 & 19:07 & 18:55 & S07E36 & B1.3 &  10988 \\ 
  66 & 2008 Mar 26 & 19:07 & 19:13 & 19:09 & S06E35 & B1.2 &  10988 \\ 
  67 & 2008 Mar 26 & 19:17 & 19:21 & 19:19 & S07E35 & B1.3 &  10988 \\ 
  68 & 2008 Mar 27 & 01:35 & 01:37 & 01:35 & S11E33 & A8.6 &  10988 \\ 
  69 & 2008 Mar 27 & 02:42 & 02:51 & 02:43 & S06E30 & B1.5 &  10988 \\ 
  70 & 2008 Mar 27 & 03:36 & 03:37 & 03:36 & S08E33 & no data &  10988 \\ 
  71 & 2008 Mar 27 & 05:17 & 05:19 & 05:18 & S07E29 & A8.9 &  10988 \\ 
  72 & 2008 Mar 27 & 11:10 & 11:13 & 11:10 & S06E26 & A8.1 &  10988 \\ 
  73 & 2008 Mar 27 & 15:50 & 15:53 & 15:50 & S05W03 & B1.3 &  10987 \\ 
  74 & 2008 Mar 27 & 16:19 & 16:20 & 16:19 & S06W03 & B1.5 &  10987 \\ 
  75 & 2008 Mar 28 & 01:11 & 01:15 & 01:11 & S05W07 & A6.5 &  10987 \\ 
  76 & 2008 Mar 28 & 09:39 & 09:48 & 09:44 & S10W11 & B1.6 &  10987 \\ 
  77 & 2008 Mar 28 & 10:44 & 10:46 & 10:45 & S07W12 & A7.2 &  10987 \\ 
  78 & 2008 Mar 28 & 14:01 & 14:10 & 14:03 & S14E13 & B1.2 &  10988 \\ 
  79 & 2008 Mar 28 & 15:27 & 15:28 & 15:28 & S10E40 & A5.8 &  10989 \\ 
  80 & 2008 Mar 30 & 07:55 & 07:56 & 07:55 & S07W39 & A5.6 &  10987 \\ 
  81 & 2008 Mar 30 & 17:32 & 17:36 & 17:36 & S07W44 & B1.3 &  10987 \\ 
  82 & 2008 Mar 30 & 20:33 & 20:36 & 20:33 & S07W48 & A8.5 &  10987 \\ 
  83 & 2008 Mar 30 & 23:25 & 23:29 & 23:26 & S07W49 & A8.3 &  10987 \\ 
  84 & 2008 Mar 31 & 06:05 & 06:06 & 06:05 & S07W54 & A8.3 &  10987 \\ 
  85 & 2008 Mar 31 & 06:32 & 06:36 & 06:33 & S08W23 & A5.5 &  10988 \\ 
  86 & 2008 Apr 01 & 00:15 & 00:16 & 00:16 & S11W05 & A7.9 &  10989 \\ 
  87 & 2008 Apr 01 & 20:15 & 20:16 & 20:15 & S11W15 & A6.5 &  10989 \\ 
  88 & 2008 Apr 01 & 20:18 & 20:22 & 20:20 & S10W15 & B1.0 &  10989 \\ 
  89 & 2008 Apr 02 & 03:09 & 03:15 & 03:10 & S09W49 & A4.3 &  10988 \\ 
  90 & 2008 Apr 02 & 12:34 & 12:38 & 12:36 & S09W55 & A8.6 &  10988 \\ 
  91 & 2008 Apr 02 & 15:44 & 15:46 & 15:45 & S08W56 & A5.1 &  10988 \\ 
  92 & 2008 Apr 02 & 16:51 & 16:58 & 16:51 & S08W61 & B1.0 &  10988 \\ 
  93 & 2008 Apr 02 & 20:03 & 20:13 & 20:06 & S16W26 & B1.1 &  10989 \\ 
  94 & 2008 Apr 02 & 22:30 & 22:31 & 22:30 & S09W61 & A6.5 &  10988 \\ 
  95 & 2008 Apr 02 & 23:40 & 23:53 & 23:43 & S09W62 & B2.0 &  10988 \\ 
  96 & 2008 Apr 03 & 01:12 & 01:50 & 01:21 & S11W32 & C1.2 &  10989 \\ 
\end{tabular}
\end{table}

\setcounter{table}{1} 

\begin{table}
\caption{Continued from previous page; see above for more information.}
\label{tab:christe_cont_again}
\begin{tabular}{rlcccccl}
  \hline                   
\# & Date & Start & Stop & Peak & Coordinates & GOES Class & Source AR \\
  \hline
  97 & 2008 Apr 03 & 04:56 & 04:57 & 04:57 & S10W64 & A8.1 &  10988 \\ 
  98 & 2008 Apr 03 & 05:55 & 06:02 & 05:57 & S09W65 & B1.2 &  10988 \\ 
  99 & 2008 Apr 03 & 06:02 & 06:10 & 06:07 & S09W65 & A9.9 &  10988 \\ 
 100 & 2008 Apr 03 & 07:35 & 07:37 & 07:36 & S10W63 & A4.2 &  10988 \\ 
 101 & 2008 Apr 03 & 20:00 & 20:11 & 20:04 & S06W77 & no data &  10988 \\ 
 102 & 2008 Apr 03 & 20:15 & 20:20 & 20:17 & S13W41 & no data &  10989 \\ 
 103 & 2008 Apr 03 & 20:20 & 20:28 & 20:23 & S13W42 & no data &  10989 \\ 
 104 & 2008 Apr 03 & 22:09 & 22:14 & 22:11 & S06W75 & A8.7 &  10988 \\ 
 105 & 2008 Apr 03 & 22:22 & 22:24 & 22:22 & S06W75 & A5.8 &  10988 \\ 
 106 & 2008 Apr 05 & 03:13 & 03:17 & 03:13 & S13W59 & A5.1 &  10989 \\ 
 107 & 2008 Apr 05 & 05:33 & 05:37 & 05:33 & S13W61 & A6.4 &  10989 \\ 
 108 & 2008 Apr 05 & 11:02 & 11:16 & 11:05 & S12W64 & B1.3 &  10989 \\ 
  \hline
\end{tabular}
\end{table}

In Table \ref{tab:flareobs}, we summarize flare activity for each WHI
AR separately for two time intervals: first, over the whole time range
from appearance over the east limb to its disappearance around the
west limb; and second, during the limited time interval that the AR
was tracked.  Not counting RHESSI flares for which a peak in the GOES
lightcurve could not be automatically isolated, we found that ARs
10987, 10988, and 10999 produced 30, 44, and 28 flares, respectively,
as they crossed the disk, and 9, 11, and 5 flares, respectively, while
they were tracked across the central disk.  \citet{Abramenko2005} used
a weighted, time-average of each AR's flare X-ray fluxes observed by
GOES as a ``flare index'' to quantify AR flare activity; consistent
with the definition of GOES flare classes, a power of 10 difference in
weighting was used between flare classes.  Given such a weighting, the
M- and C-class flares produced by AR 10989 imply that its average
flare flux greatly exceeds that of the other two ARs during either of the
intervals in Table \ref{tab:flareobs}.

\begin{table}
\caption{Observed Flares and CMEs from WHI ARs, from limb to limb (top
  three rows), and while tracked (bottom three rows).  A, B, C, and M
  denote GOES flare classes. (RHESSI flares during GOES data gaps are
  not included in these tallies.)  CMEs within a few hours before or
  after the start or end times shown are included in parentheses
  before or after the frequency during the interval, respectively.
  The M- and C-class flares produced by AR 10989 imply that its flare
  index, a measure of AR flare productivity (Abramenko, 2005),
  exceeds those of ARs 10987 and 10988.}
\label{tab:flareobs}
\begin{tabular}{lllcccccc} 
  \hline                   
AR & Start Obs. & End Obs. & A & B & C & M & Tot. Flares & CMEs \\
  \hline
10987 & Mar 23, 23:39UT & Apr 05, 11:17UT & 11 & 19 &    &   & 30 & 0\,(1) \\ %
10988 & Mar 23, 18:51UT & Apr 05, 11:17UT & 24 & 20 &    &   & 44 & (1)\,4  \\ %
10989 & Mar 24, 19:09UT & Apr 05, 11:17UT &  5 & 20 &  2 & 1 & 28 & 8\,(1) \\ 
  \hline
10987 & Mar 24, 14:23UT & Mar 30, 04:47UT &  2 &  7 &    &   & 9 & 0 \\ %
10988 & Mar 26, 12:48UT & Apr 01, 06:23UT &  4 &  7 &    &   & 11 & 0 \\ %
10989 & Mar 28, 22:23UT & Apr 03, 14:23UT &  2 &  2 &  1 &   & 5 & (1)\,2\,(1) \\ 
  \hline
\end{tabular}
\end{table}

\section{Analysis of Magnetic Structure and Evolution}
\label{sec:methods}

We quantitatively analyzed magnetic evolution in these three ARs in
several ways, which we present roughly in order of decreasing length
scale: starting from whole-active-region measures of magnetic
structure and evolution, we consider evolution on progressively
smaller spatial scales, down to individual MDI pixels.

\subsection{Large Scale Structure and Evolution}
\label{subsec:global} 

Barnes and Leka (2008), Leka and Barnes (2007), and \citet{Welsch2009}
have noted that total unsigned radial magnetic flux $[\Phi]$ in an AR
is strongly correlated with its flare productivity.

\begin{figure}
  \centerline{\includegraphics[width=0.75\textwidth]{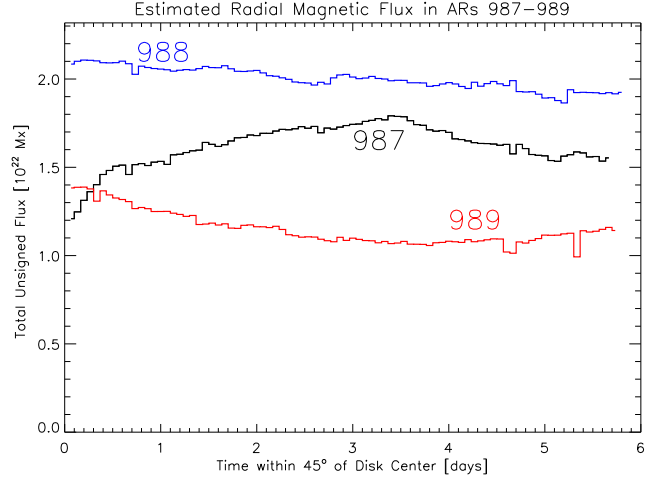}}
  \caption{Total unsigned flux $[\Phi]$ {\it versus} time in the three WHI ARs, as each
    crossed the central solar disk.  Dips arise from lower noise levels
    in magnetograms averaged over five measurements.  Larger values of 
  $\Phi$ have been associated with greater flare activity.}
  \label{fig:flux_vs_time}
\end{figure}

We calculated the total unsigned estimated radial flux, $[\Phi = \int
  \mathrm{d}A \, |B_R|]$, in each of the WHI ARs over the time
interval each was tracked, compensating for distortion in pixel areas
due to the Mercator projection.  These unsigned fluxes are plotted in
Figure \ref{fig:flux_vs_time}.  The average values of unsigned flux
for ARs 10987, 10988, and 10989 were 1.6, 2.0, and 1.2 $\times 10^{22}$
Mx, respectively, over the interval when each was tracked.  To
minimize the effects of noise, only pixels with unsigned field
strength greater than 40G were included in the sums; this is
approximately three times our assumed noise level of 14G.  Assuming
the uncorrelated, uniform per-pixel noise level on flux of $\sigma_0 =
14$G$(\Delta x)^2$, where $\Delta x$ is the width of MDI full-disk
pixels, a formal estimate of the error in total flux would be
$\sigma_{\rm sum} = \sqrt{N_{\rm > 40G}} \sigma_0$; this is typically
$< 10^{20}$ Mx, which is not visible on these plots.  Dips in the
plots arise from lower noise levels in magnetograms averaged over five
measurements, and give a better estimate of the noise level.  Also,
the magnetic field is highly correlated from frame to frame at 96
minute cadences \citep{Welsch2009}, implying frame-to-frame variations
in total flux also indicate noise levels.  Note that MDI has a known
problem with saturation in strong-field regions \cite{Liu2007b}, such
that flux values reported here might be lower than the actual flux
\cite{Wang2009}.

Based upon total unsigned flux alone, AR 10988 should have been the
most active, and AR 10989 the least.

\citet{Schrijver2007} argued that strong magnetic fields concentrated
near polarity inversion lines (PILs) of the photospheric LOS magnetic
field are closely associated with the occurrence of large flares.  He
developed a method to quantify the total unsigned magnetic flux $[R]$
near strong-field polarity inversion lines (SPILs) of ARs, by
generating a weighting map of proximity to SPILs and summing the
product of the weighting map and unsigned LOS magnetic flux.  Because
strong fields along LOS PILs are correlated with strong gradients in
LOS fields across PILs \citep{Falconer2003}, the $R$ parameter is
quantitatively related to the length of strong-gradient PILs, which
has been found by \citet{Falconer2003} and \citet{Falconer2006} to be
  associated with CMEs.  \citet{Barnes2008} and \citet{Welsch2009}
  also found $R$ to be associated with flare activity.

\begin{figure}
  \centerline{\includegraphics[width=0.75\textwidth]{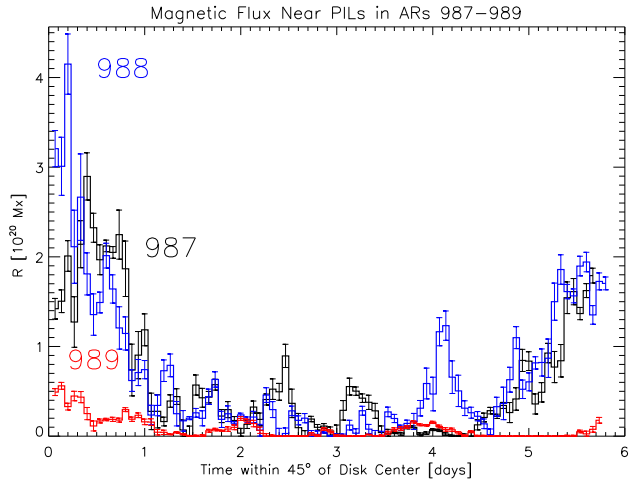}}
  \caption{Total unsigned flux $[R]$ near strong-field polarity
    inversion lines (SPILs) {\it versus} time in the three WHI ARs, as each
    crossed the central solar disk.  Large values of $R$ have been
    associated with flare activity.}
  \label{fig:r_vs_time}
\end{figure}

Emulating Schrijver's approach here, we computed $R$ values for each
of the WHI ARs as they crossed the disk, choosing (as he did) a
strong-field threshold of 150G and FWHM of 15Mm in the Gaussian used
in the convolution to compute the SPIL weighting map (see
\citealt{Welsch2008b} for more details of the procedure).  Our results
are plotted in Figure \ref{fig:r_vs_time}.  Note that our values of
$R$ are in units of Mx; Schrijver's were in units of G, summed over
the weighting map.  The MDI disk-center pixel area of $\simeq 2.2
\times 10^{16}$ cm$^2$ is an approximate conversion factor. Note also
that \citet{Schrijver2007} apparently used MDI Level 1.8.0
magnetograms, while we use Level 1.8.2 magnetograms (see
\url{http://soi.stanford.edu/magnetic/Lev1.8/}). Since absolute flux
densities in the latter are higher than in the former by a spatially
varying factor of approximately 1.6, our $R$ values cannot easily be
compared with Schrijver's.  As with estimating total unsigned flux,
uncertainties in $R$ computed assuming an uncorrelated, uniform
per-pixel noise level $[\sigma_0]$ are are too small to be visible on
the plot.  Instead, we have crudely estimated uncertainties from the
$R$ values themselves by computing the standard error in the mean with
a five-point boxcar window.  This simplistic approach is reasonable
because, as noted by Welsch \etal (2009), changes
in magnetic fields are relatively minor over a few hours, so
successive measurements of $R$ can be interpreted as repeated,
independent measurements of nearly the same physical value.

As with total unsigned flux, expectations of flare activity based upon
$R$ values would suggest AR 10988 should be most active, and AR 10989
should be least active.

\subsection{Intermediate-Scale Evolution: Feature Tracking}
\label{subsec:feature}

Automated tracking of ``features'' in magnetogram sequences has been
used to understand processes governing evolution of the photospheric
magnetic field, including flux emergence and dispersal ({\it e.g.}
\citealt{DeForest2007} and references therein).  Features in
magnetograms have been identified in several ways, for instance by:
``clumping'' collections of contiguous, like-polarity, above-threshold
pixels ({\it e.g.} \citealt{Parnell2009}); ``downhill'' segmentation,
which identifies distinct ``hilltops'' in absolute field strength
({\it e.g.} \citealt{Welsch2003}); and curvature, identifying the
convex cores of hilltops in absolute field strength ({\it e.g.}
\citealt{Hagenaar1999}).  Tracking involves the association of
identified features between successive frames.

Feature tracking can be used to identify episodes of flux
cancellation, in which closely spaced, opposite-polarity features
simultaneously appear to lose flux \citep{Livi1985}.  Since flux
cancellation has been observed in prominence formation
\cite{Martin1998}, shown to increase magnetic free energy in some
circumstances \cite{Welsch2006}, and proposed as a CME initiation
mechanism ({\it e.g.} \citealt{Linker2003}), we tracked downhill
features in magnetogram sequences of the WHI ARs, to investigate
cancellation rates in each AR.

\begin{figure}
  \centerline{\includegraphics[width=0.95\textwidth]{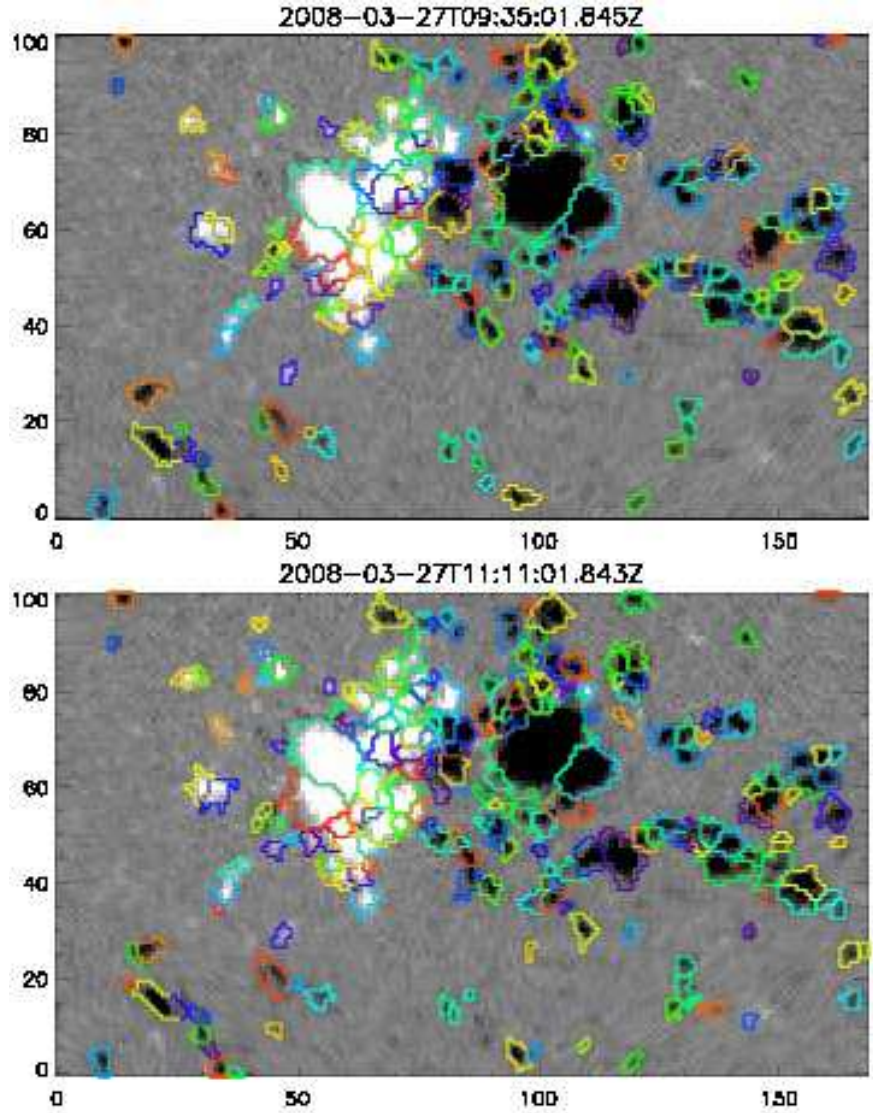}}
  \caption{The outlines of features --- ``hills'' in absolute field
    strength --- identified in two successive magnetograms from AR
    10987 are plotted over grayscale images of the magnetic field
    (white is positive polarity, black negative; saturation is set at
    $\pm 250$G). The outlines are color coded by feature label, and
    features have been matched between these magnetograms.
    Different-color outlines imply distinct features were identified;
    same-color outlines generally imply matching features between
    frames.}
  \label{fig:feature_combo}
\end{figure}

Using the YAFTA feature-tracking code (\citealt{Welsch2003},
\citealt{DeForest2007}; software and documentation are online at
\url{http://solarmuri.ssl.berkeley.edu/~welsch/public/software/YAFTA/})
we only included pixels above 35G (a 2.5$\sigma$ threshold) in
features, and required each feature to have at least four pixels, and
at least one 70G pixel (a 5$\sigma$ peak threshold).  We also required
features to either persist for at least four frames, or interact with
(fragment from, or merge with) a feature that persisted for four
frames.  The outlines of features identified in two successive
magnetograms from AR 10987 are plotted over grayscale images of the
magnetic field in Figure \ref{fig:feature_combo}. The outlines are
color coded by feature label, and features have been matched between
these magnetograms.  Different-color outlines imply distinct features
were identified; same-color outlines generally imply matching features
between frames, although the color palette used is limited, so
spurious color matches are possible.  While substantial evolution has
occurred over the 96 minutes between these magnetograms, 90\% of
features in the first magnetogram were identified in the second, which
is a typical survival rate.

Having tracked the three ARs' magnetogram sequences, we then attempted
to quantify canceling features in each.  We first found
opposite-polarity features in close proximity by looping over positive
features, applying IDL's \url{dilate} function, with a $(3 \times 3)$
structuring element, to each feature's pixels; and seeking instances
of overlap between the dilated feature's pixels and pixels of negative
features.  This essentially searched for negative features in
nearest-neighbor pixels on the periphery of each positive
feature. Hence, our definition of close proximity for defining
cancellation is MDI's pixel size, $\simeq 2'' \simeq 1.4$ Mm.  For a
pair of close features to have partially cancelled over a time step,
we further required the features' centers of flux to approach, and the
flux of both features to decrease --- and by doing so, we have
implicitly assumed that neither feature disappears completely over a
cancellation step.

This approach is subject to many sources of error.  Thresholds imposed
on field strengths in features imply that some magnetic flux is not
included in any feature.  Features' positions jitter, and their fluxes
fluctuate, both from noise and true evolution.  Further, along the
polarity-inversion lines where canceling features are typically
identified, each feature often interacts with several neighboring
like-polarity features, and possibly other opposite-polarity features,
perhaps exchanging flux or canceling with those features, too.
Hence, there are many essentially random sources of changes in
features' fluxes and locations.  With higher-cadence data, additional
constraints might be imposed, {\it e.g.} meeting our cancellation criteria
over several time steps.  However, a more-detailed analysis of
cancellation in higher-cadence data lies outside the purview of this
study.  Further, we are primarily interested in comparing cancelled
flux between the WHI ARs, not establishing absolute amounts of flux
cancelled.

Recognizing that our approach to quantifying cancelled flux is subject
to large uncertainties, we computed estimates in three ways: {\it i)}
first, for each pair of canceling features identified, we took the
{\em minimum} flux loss from the pair as the cancelled flux in that
event, and summed these fluxes from all cancellation events; {\em ii)}
we limited our analysis to pairs of features that were observed to
cancel over at least two steps, and took the {\em average} flux losses
from such events; and {\em iii)} again restricting our attention to
features that were observed to cancel over at least two steps, we
summed the {\em minimum} flux lost from each pair in each event.  The
third approach is presumably the most restrictive method of estimating
cancelled flux.

For each AR, our tabulation of the total number of cancellation
events, the number of multi-step cancellation events, and estimates
for cancelled flux are listed in Table \ref{tab:cancel}.  Variations
between the estimates for a single AR provide a measure of the
uncertainties, although as noted above, the method used for the third
estimate is the most restrictive.  If our identification of canceling
features were essentially random, and mean feature fluxes in all three
ARs were the same, then our estimates of cancellation events and
cancelled fluxes would follow the ratio of unsigned magnetic fluxes in
the ARs, 1.4:1.7:1 (see Section \ref{subsec:global} above).  While the
numbers in each column do follow the ordering of these ratios, there
is an excess of both cancellation events and cancelled flux in AR
10988 beyond that expected from variations in AR flux alone.

As with both total unsigned flux and unsigned flux $[R]$ near SPILs, our
analysis of flux cancellation in the WHI ARs suggests that AR 10988
should have been the most active, and AR 10989 the least.

\begin{table}
\caption{Estimates of flux cancelled in the WHI ARs, computed three
  ways (see text); cancelled flux values are in units of [$10^{20}$ Mx].}
\label{tab:cancel}
\begin{tabular}{cccccc}
  \hline                   
AR & \# Events & \# Multi-step & All events, & Multi-step,  & Multi-step, \\
 & & 
& $\sum$ min$(\delta \Phi)$ 
& $\sum$ avg$(\delta \Phi)$ 
& $\sum$ min$(\delta \Phi)$  \\
  \hline
10987 &  58 & 10  & 5.0 & 4.0 & 1.9 \\
10988 & 110 & 21  & 8.0 & 8.1 & 3.4 \\
10989 &  42 &  5  & 2.9 & 1.6 & 0.8 \\
  \hline
\end{tabular}
\end{table}

\subsection{Pixel-Scale Evolution}
\label{subsec:pixel}

\subsubsection{Estimating Velocities}
\label{subsubsec:flct}

According to Faraday's law, evolution of the radial magnetic field
at the photosphere is governed by the curl of the electric field there,
\be \partial_t B_r = -c (\nabla \times \evec)_r
~. \label{eqn:faraday} \ee
Assuming that the electric field is ideal (equivalently, that the
plasma's conductivity is infinite) implies $\evec = -(\vvec \times
\bvec)/c$, so evolution of the radial magnetic field can be related to
horizontal variations in the velocity and magnetic fields by the ideal
induction equation,
\bea \partial_t B_r 
&=& [\nabla \times (\vvec \times \bvec)]_r \\
&=& - \nabla_h \cdot (\vvec_h B_r - v_r \bvec_h) 
~. \label{eqn:induction} \eea
The radial flux $S_r$ of magnetic energy across the photosphere (the
Poynting flux) depends upon $\evec$, and in the ideal approximation
$\vvec$,
\be S_r = c[\evec \times \bvec]_r/4\pi 
= [v_r B_h^2 - (\vvec_h \cdot \bvec_h) B_r]/4\pi
, \label{eqn:poynting0} \ee
as does the rate of change of relative magnetic helicity
\cite{Berger1984} in the corona $\mathrm{d}H/\mathrm{d}t$ due to the
helicity flux across the photosphere,
\be \frac{\mathrm{d}H}{\mathrm{d}t} 
= 2 \int \mathrm{d}A \, 
[ ( \avec_h^P \cdot \bvec_h) v_r 
- ( \avec_h^P \cdot \vvec_h) B_r]
, \label{eqn:berger0} \ee
where: the integration runs over the photosphere in the region of
non-zero $\vvec$ and $\bvec$; $\avec^P$ is the vector potential of the
current-free (and therefore curl-free, or ``potential'') magnetic
field $\bvec^P$ that matches the observed photospheric radial field
$B_r$, {\em i.e.} $\nabla_h \times \avec_h^P = B_r^P$; and $A_r^P = 0 =
\nabla_h \cdot \avec_h^P$.

\citet{Demoulin2003} argued that the ``footpoints'' of magnetic fields
anchored in the photosphere appear to move in magnetograms with an
apparent footpoint velocity $\Uvec$, which is related to the plasma
velocity $\vvec$ by
\be \Uvec \equiv \vvec_h - (v_r/B_r) \bvec_h 
~. \label{eqn:dnb} \ee
They further argued that tracking methods applied to magnetograms,
such as local correlation tracking ({\em e.g.} \citealt{Chae2001}), would
estimate $\Uvec$, not $\vvec_h$.  \citet{Welsch2006} referred to
$\Uvec$ as the flux transport velocity; $\Uvec B_r$ has units of a
flux transport rate (maxwells per unit length per unit time).  Note
that Equation (\ref{eqn:dnb}) is a matter of definition, and so is
made without approximation.  In terms of $\Uvec$, Equation
(\ref{eqn:induction}) can be written in a form analogous to the
continuity equation,
\be \partial_t B_r + \nabla \cdot (\Uvec B_r) = 0
~. \label{eqn:induct_dnb} \ee

The statement that tracking methods will accurately estimate $\Uvec$,
however, is a testable assertion; \citet{Schuck2008} argued, in
fact, that some ``optical flow'' methods are relatively insensitive to
radial flows, so will return a biased estimate $\tilde \Uvec$ of the
true $\Uvec$.  Tests of several velocity estimation methods conducted
by \citet{Welsch2007} using synthetic data from MHD simulations, in
which the true velocities were known, demonstrated that such methods
were far from perfect, but that their flow estimates were
highly correlated with the true flows.

In terms of the flux transport velocity, the Poynting and helicity
fluxes in Equations (\ref{eqn:poynting0}) and (\ref{eqn:berger0})
become
\bea
S_r &=& -(\bvec_h \cdot \Uvec_h) B_r / 4\pi \label{eqn:poynting1} \\
\frac{\mathrm{d}H}{\mathrm{d}t} &=& = - 2 \int \mathrm{d}A \, 
( \avec_h^P \cdot \Uvec_h) B_r 
~. \label{eqn:berger1} \eea
Note that estimating the Poynting flux requires knowledge of the
horizontal magnetic field.

The flux transport velocity can be estimated from magnetogram sequences, 
with finite cadence $\Delta t$ and pixel size $\Delta x$, by applying a 
finite difference approximation to Equation (\ref{eqn:induction}),
\be \frac{\Delta B_z}{\Delta t} + 
\frac{{\bf \Delta} \cdot (\Uvec B_r)} {\Delta x} = 0 
~, \label{eqn:induct_fd} \ee
where ${\bf \Delta}$ is the spatial finite-difference operator.
If a typical flow speed is $v_0$, then the pixel-crossing time is
$\tau_0 = v_0 \, \Delta x$.  If $\Delta t < \tau_0$, then Equation
(\ref{eqn:induct_fd}) approximates Equation (\ref{eqn:induct_dnb});
if, however, $\Delta t > \tau_0$, the approximation can fail.  This is
analogous to the Courant-Friedrichs-Lewy (CFL) condition required for
numerical stability in computational fluid dynamics.

\citet{Welsch2009} used Equation (\ref{eqn:induct_fd}) as the basis
for applying two tracking methods, Fourier local correlation tracking
(FLCT: \citealt{Welsch2004},\citealt{Fisher2008}) and the differential
affine velocity estimator (DAVE: \citealt{Schuck2006}), to sequences
of 96-minute MDI full-disk magnetograms (similar to the current data,
but Level 1.8.0 instead of Level 1.8.2).  Both methods employed
similar windowing parameters (eight and nine pixels for FLCT and DAVE,
respectively) to localize the magnetic data used in estimating the
flow at a given pixel, which probably has the effect of averaging over
smaller-scale flows.  Correlation coefficients between the
field-weighted FLCT and DAVE flows' $x-$ and $y-$components were $>
0.7$ over the set of all the flows computed by Welsch \etal (2009),
implying consistency between flow estimates using distinct methods.
Mean and median speeds for FLCT were $\simeq 65$ m sec$^{-1}$, and for
DAVE were $\simeq 105$ m sec$^{-1}$; evidently, DAVE speeds are higher
than those estimated with FLCT.  Given the $\simeq 1.4$ Mm disk-center
width of MDI full-disk pixels, these speeds imply pixel-crossing times
of $\sim 4-6$ hr.  \citet{Welsch2009} also found that flow directions
remained significantly correlated over similar time scales.  Both of
these facts imply that 96-minute cadence, full-disk MDI data can be
used with Equation (\ref{eqn:induct_fd}) as a valid approximation of
Equation (\ref{eqn:induct_dnb}).

We applied FLCT to the 96-minute cadence MDI full-disk magnetograms
within 23 Mar 2008, 00:00:00UT - 4 Apr 2008, 00:00:00UT from the WHI
interval, with a windowing parameter of eight pixels, an absolute
field-strength threshold for tracking of 50 G, and a low-pass spatial
wavenumber roll-off parameter of 0.25 (see \citealt{Fisher2008}).
Pixels out to 45$^\circ$ from disk center were tracked.  This choice
of windowing parameter corresponds to $\simeq 10$Mm, a scale much larger
than granules, but slightly smaller than supergranules
\citep{Hagenaar1997}.

\begin{figure}
  \centerline{\includegraphics[width=0.95\textwidth,clip=true]{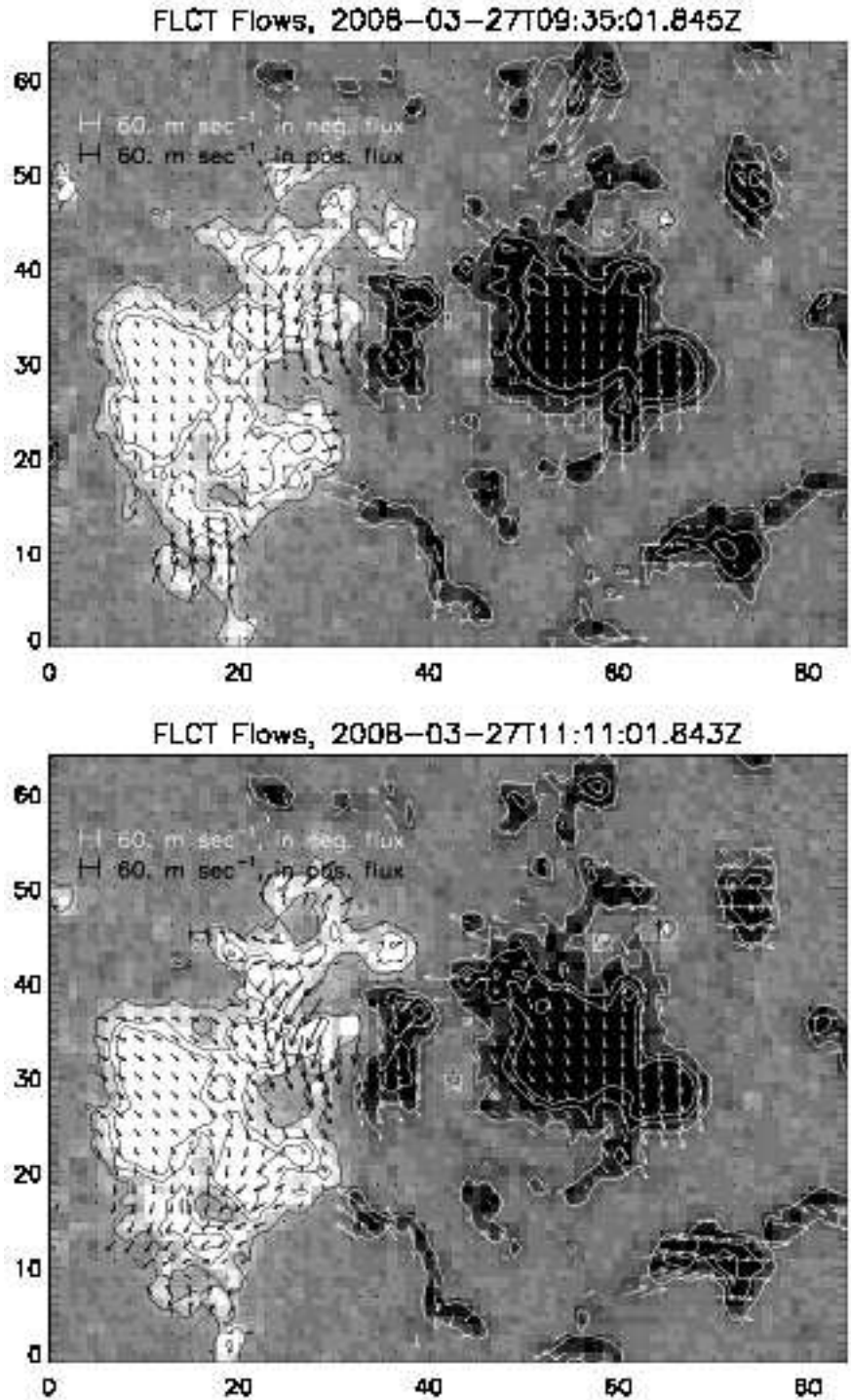}}
  \caption{Two successive FLCT flow maps from the central,
    strong-field portion of AR 10987, from the same magnetograms shown
    in Figure \ref{fig:feature_combo}.  Background grayscale shows
    estimated radial field strength, white showing positive flux,
    black negative, with saturation set at $\pm 250$G.  White/black
    contours correspond to $\pm100, \pm300,$ and $\pm500$G.  For
    clarity, only every other flow vector is plotted. Some evolution
    in the flow pattern can be seen; but flows in many regions appear
    generally similar at both times.}
  \label{fig:flct_combo}
\end{figure}

Data cubes covering each of the WHI ARs' magnetic fields and their
corresponding flows were then extracted from the full-disk
magnetograms and inner-disk flow maps for the time intervals listed in
the bottom three rows of Table \ref{tab:flareobs}.  Figure
\ref{fig:flct_combo} shows two successive FLCT flow maps from the
central, strong-field portion of AR 10987, from the same magnetograms
shown in Figure \ref{fig:feature_combo}.  Some evolution in the flow
pattern can be seen over the 96 minutes between the flow maps; but
flows in many regions appear qualitatively similar at both times.

\subsubsection{Proxy Poynting Flux}
\label{subsubsec:proxy}

\citet{Welsch2009} analyzed relationships of flare activity with
photospheric magnetic field and flow properties in a sample of a few
dozen ARs.  Of several quantities that they investigated, they found
Schrijver's $R$ value and a quantity proportional to $\sum (u B_R^2)$,
where $u = |\Uvec|$, and the sum runs over the AR magnetograms, to be
most strongly associated with flare activity.  Because $u B_R^2$ has
the same dimensions as the Poynting flux of magnetic energy (erg
cm$^{-2}$ sec$^{-1}$), \citet{Welsch2009} called this quantity the
proxy Poynting flux.  For consistency with Equation
(\ref{eqn:poynting1}), we define the proxy Poynting flux here to be
$S_R \equiv (u B_R^2)/4 \pi.$ (Note that this definition differs from
that of Welsch \etal (2009), who defined $S_R$
to include a sum over pixels --- a total flux, instead of a flux
density --- and did not include $4\pi$.)  \citet{Li2010} estimated the
proxy Poynting flux in AR 8038 around the time of the well-known 12
May 1997 flare/CME, and found the cumulative flux reached $\simeq 10^{32}$
erg over the four-day interval before the eruption.

What is the physical basis for the relationship between the proxy
Poynting flux and flare activity?  Leaving aside the likelihood that
estimates of $B_r$ and $\Uvec$ are imperfect, in principle, $S_R$
corresponds to part of the horizontal Poynting flux $[S_h]$
(additional contributions arise from terms containing $\bvec_h$).  The
magnetic energy that powers flares and CMEs enters the corona from the
solar interior, as an outward, radial Poynting flux $[S_r]$.  It is
plausible that there should be a correlation between $S_r$ and
flare/CME activity, although the corona can store magnetic energy, so
there might typically be a latency between the introduction of
magnetic energy into the corona and its release in flares/CMEs.  For
instance, \citet{Longcope2005} and \citet{Schrijver2005} report
latency times associated with active-region magnetic fields of $\sim
24$ hours. (And in quiet-Sun regions, filaments/ prominences, which
are interpreted as current-carrying [non-potential] magnetic field
structures, can persist in the corona for weeks.)  It is also
plausible that the radial Poynting flux $[S_r]$ is significantly
correlated with $[S_h]$ and/or the proxy flux $[S_R]$, but no
quantitative theoretical or observational characterization of such a
correlation has been presented.

\begin{figure}
  \centerline{\includegraphics[clip=true]{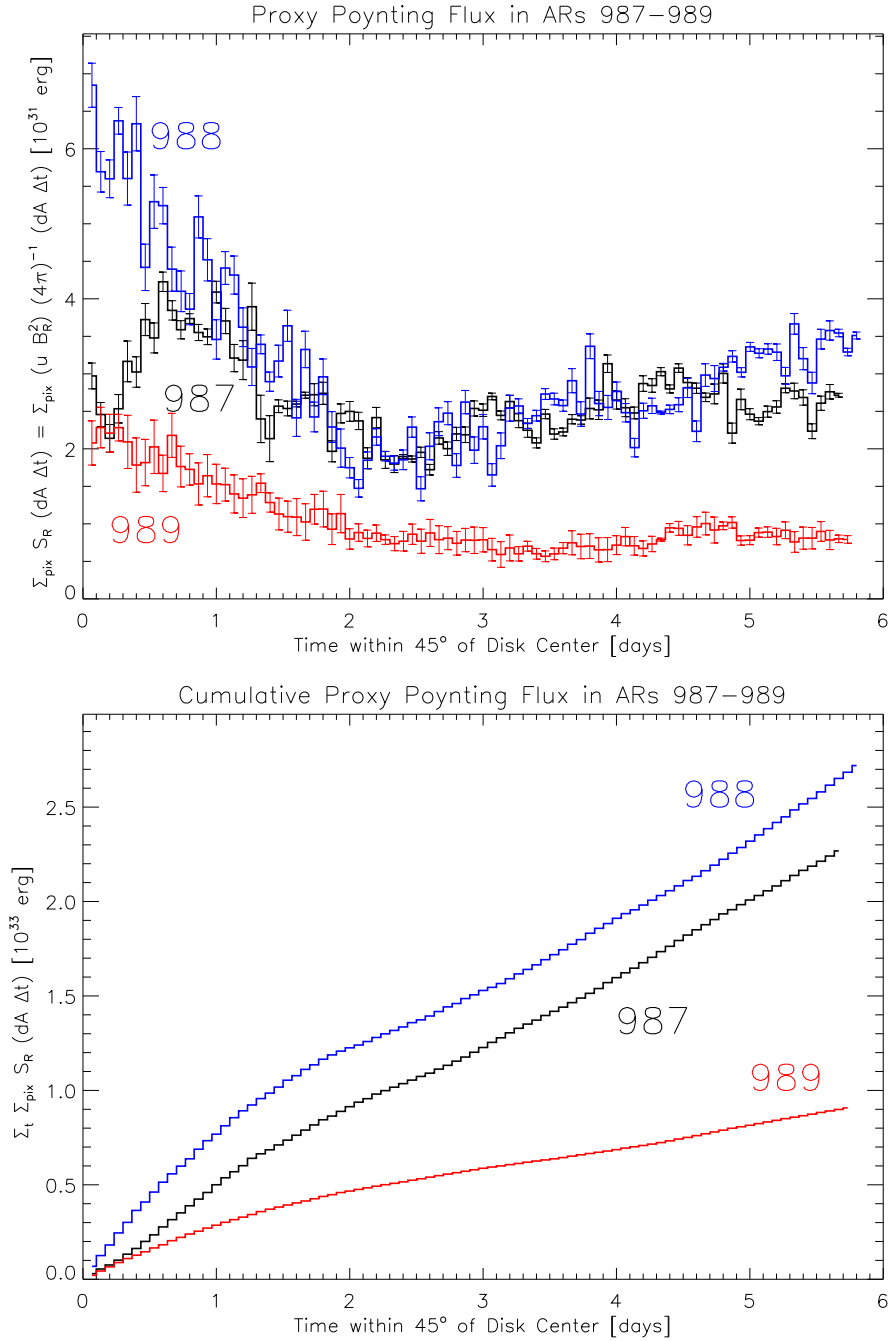}}
  \caption{The proxy Poynting flux $[S_R]$ multiplied by pixel area
    $[\Delta A]$ and the time between magnetograms $[\Delta t]$ (top) and its
    sum over time (bottom) are both plotted {\it versus} time for the three WHI
    ARs, as each crossed the central solar disk.  Error bars for the
    cumulative plot would be about as large as vertical step sizes.
    Large values of $\sum_{\rm pix} S_R$ have been associated with
    flare activity.  The sum over $S_R$ can provide an
    order-of-magnitude estimate of the magnetic-energy transport
    rates.}
  \label{fig:sr_vs_time}
\end{figure}

Using the FLCT flows we estimated for each WHI AR, we computed the
proxy Poynting flux in that AR over the time interval between each
tracked magnetogram pair.  We then multiplied by (reprojection-
corrected) pixel area $\Delta A$ and the time interval $\Delta t$ between
magnetograms, and summed over all pixels to express our results in
terms of ergs of energy transported.  Results are plotted in the top
panel of Figure \ref{fig:sr_vs_time}; in the bottom panel, we plot the
cumulative proxy Poynting fluxes over time.  As with $R$, we estimated
uncertainties in the top panel from a running-boxcar computation of
the standard error in the mean.  Assuming a combined fractional
uncertainty of 0.6 for the terms in the product $u B_R^2$ in each
pixel resulted in error bars smaller than those shown.  Summing the
uncertainties in the top panel in quadrature for the cumulative result
shown in the bottom panel yields error bars about the size of the
steps in the bottom panel.

Unlike the true Poynting flux, which can be positive or negative ({\it
  i.e.} radially outward or inward), the proxy Poynting flux is only
positive.  To the extent that the proxy Poynting flux reflects the
outward radial flux of magnetic energy, the values in these plots
suggest the order of magnitude of energy transport in the WHI ARs.

As with other magnetic parameters of the WHI ARs, values of the proxy
Poynting flux suggest AR 10988 should have been the most active, and
AR 10989 the least.

\subsubsection{Helicity Flux}
\label{subsubsec:helicity}

Magnetic helicity is approximately conserved even when magnetic
reconnection occurs, meaning that it is not readily dissipated in the
corona \citep{Berger1984c}, and can accumulate there as twisted
magnetic flux emerges from the interior \citep{Pevtsov1997}.  It is
plausible that helicity accumulation in the corona plays a key role in
triggering CMEs \cite{Low2001}.

\begin{figure}
  \centerline{\includegraphics[clip=true]{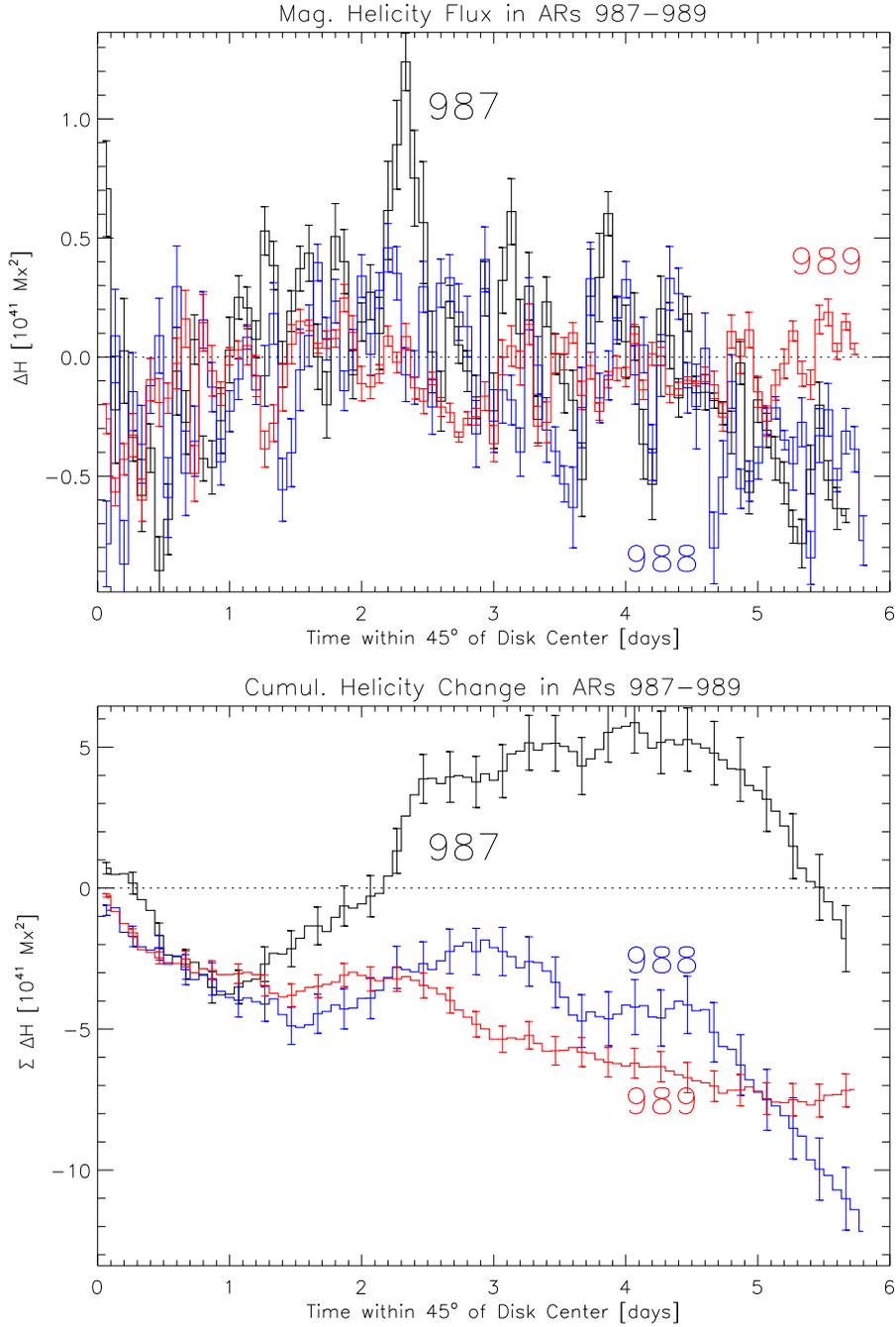}}
  \caption{Top: the differential fluxes of helicity for the WHI ARs,
    with uncertainties. Bottom: The cumulative helicity fluxes of each
    WHI AR; only every third error bar is plotted, for clarity.}
  \label{fig:dhdt_combo}
\end{figure}

Velocities inferred from tracking can also be used to estimate the
flux of magnetic helicity across the photosphere (e.g.,
\citealt{Chae2001}, \citealt{Kusano2002}, \citealt{Pariat2005}).
Using the FLCT flows we estimated for each WHI AR, we computed the
helicity flux in that AR over the time interval between each tracked
magnetogram pair, separately using {\it i)} Equation
(\ref{eqn:berger1}) with a Green's function method to calculated
$\avec^P$, and {\it ii)} the method presented by Pariat, D\'emoulin,
and Berger (2005).  Means, medians, and linear fits between the two
methods agreed to within 3\% for AR 10987, 1\% for AR 10988, and 6\%
for AR 10989.  In the top panel of Figure \ref{fig:dhdt_combo}, we
show the differential fluxes of helicity for the WHI ARs computed from
Equation (\ref{eqn:berger1}), with uncertainties again computed from a
running-boxcar calculation of the standard error in flux estimates.
(As above, formally propagated errors were too small to appear on
these plots.)  In the lower panel of Figure \ref{fig:dhdt_combo} we
plot the cumulative helicity fluxes of each WHI AR; only every third
error bar is plotted, for clarity.

\begin{figure}
  \centerline{\includegraphics[width=0.95\textwidth,clip=true]{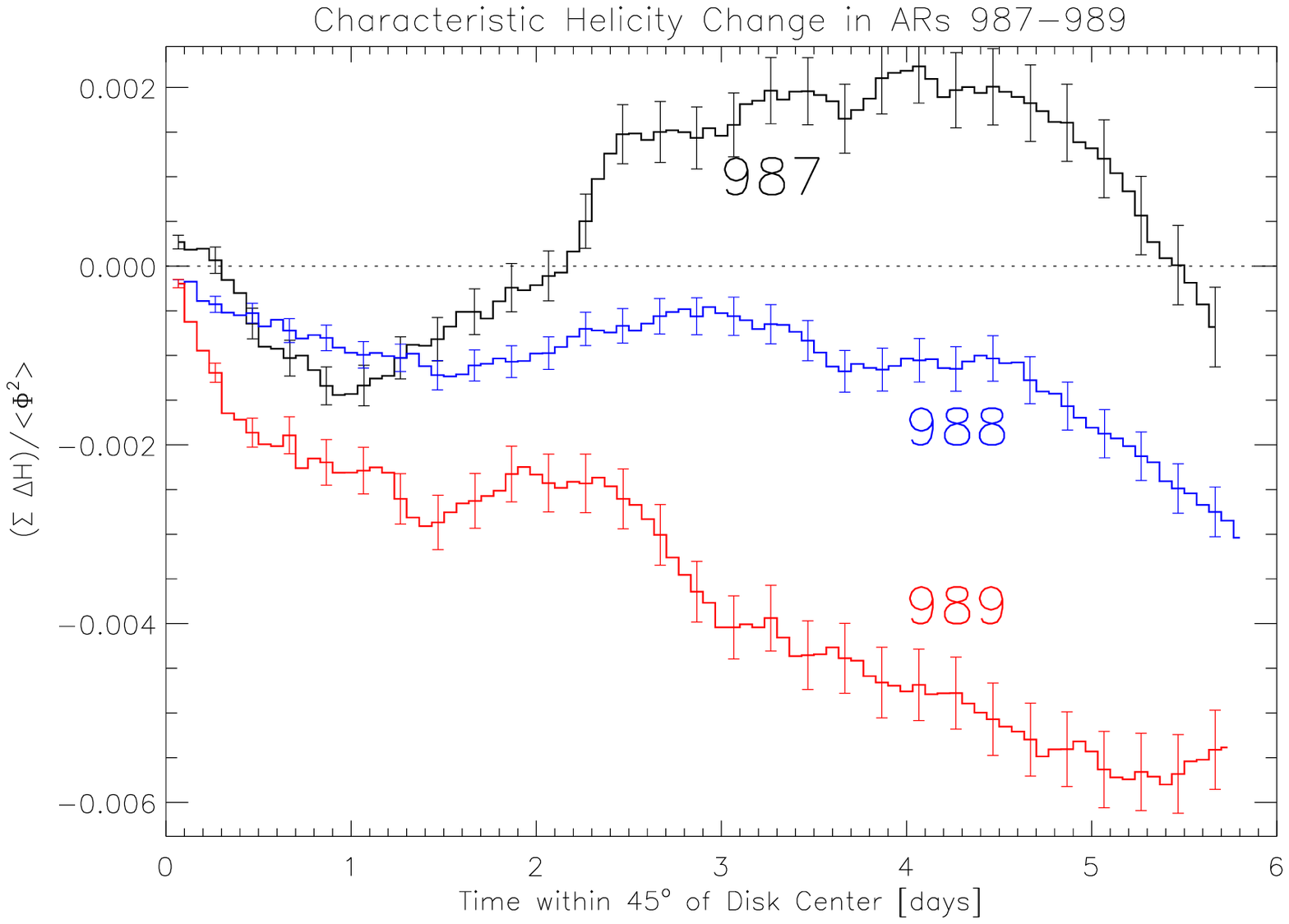}}
  \caption{The cumulative characteristic helicity fluxes of each WHI
    AR: the cumulative helicity flux, normalized by the square of
    the mean magnetic flux in each AR over the time period it was
    observed.  The unnormalized, cumulative, helicity fluxes for ARs
    10988 and 10989 are similar, but the characteristic helicity for
    AR 10989 is substantially higher than for 10988.}
  \label{fig:dhdt_char}
\end{figure}

A characteristic magnitude for helicity is the flux in the system
squared.  In Figure \ref{fig:dhdt_char}, we plot the cumulative
helicity fluxes of each WHI AR, normalized by the square of the mean
magnetic flux in each AR over the time period it was observed.  The
unnormalized, cumulative, helicity fluxes for ARs 10988 and 10989 are
similar; but since the relative magnetic fluxes of the ARs 10987,
10988, and 10989 are 1.4:1.7:1, normalizing by their squared fluxes
implies the characteristic helicity for AR 10989 is substantially
higher than for 10988.  

We have reviewed several magnetic parameters of the WHI ARs that might
be correlated with flare and CME activity.  Characteristic helicity is
the only such parameter for which AR 10989 has a larger value than the
other ARs. The characteristic helicity fluxes that we found, however,
are far below unity, implying that the helicity fluxes did not inject
large amounts of twist into the WHI ARs, relative to their magnetic
size.

The total helicity present in an AR is the combination of that present
in the AR when it emerged and the helicity injected after emergence.
By extrapolating non-linear force-free fields (NLFFFs) from a vector
magnetogram, the instantaneous helicity content of an AR can be
estimated.  \citet{Petrie2011}  have done this for the WHI ARs with
SOLIS \citep{Henney2009} data, and report characteristic helicities
(see their Table 1) of $0.0055, 0.0041,$ and $0.0017$ for ARs 10987,
10988, 10989, respectively.  Note that all of their characteristic
helicity values are on the same order as those we report, but that all
of their values are positive; and that their value of the
characteristic helicity in AR 10989 is less than those of the other
WHI ARs, undermining the hypothesis that characteristic helicity might
explain variations in flare and CME productivity. 

As an aside, we note that some of our estimates of helicity {\em
  changes} also differ those of \citet{Petrie2011} (see the right
column of their Figure 23).  They show AR 10987 with initially
positive helicity that tends to decrease with time, which only
partially agrees with our results, which show an initially negative
helicity flux followed by a larger, positive helicity flux.  Our
estimates show negative helicity steadily being added to AR 10988 over
time, but \citet{Petrie2011} found increasingly positive helicity over
one time interval.  Finally, the estimates of helicity changes in AR
10989 by \citet{Petrie2011} are relatively weak, and of mixed sign;
but our results show a steady flux of negative helicity.

\subsection{Inductive Flow Estimation}
\label{subsec:induct}

Flows can be estimated from the induction equation alone, by first 
applying a Helmholtz decomposition to the horizontal electric field,
\be c \evec_h = -\nabla_h \phi + \nabla_h \times \chi \uvec{r} 
~, \ee
and using this result in Equation (\ref{eqn:faraday}), resulting
in a Poisson's equation for $\chi$,
\be  \partial_t B_r = \nabla_h^2 \chi 
~. \label{eqn:poisson} \ee
If the electric field is assumed to be ideal, then $c \evec_h = -\Uvec
\times B_r \uvec{r}$, and a ``Poisson flow'' can be derived from $\chi$,
\be \Uvec B_r = -\nabla_h \chi ~. \label{eqn:poisson_flow} \ee
We solved the finite-difference approximation to Equation
(\ref{eqn:poisson}) for the WHI AR magnetogram sequences using Fast
Fourier Transforms (FFTs), and then derived Poisson flows.  Mean and
median Poisson flow speeds for the pair of magnetograms shown in
Figure \ref{fig:flct_combo} were 91 and 63 m sec$^{-1}$, respectively,
slow enough to be consistent with the finite-difference approximation.

One problem with this approach is that fluctuations in individual
pixels due purely to noise or observational artifacts are ascribed to
actual electric fields or flows.  In contrast, tracking methods such as
FLCT and DAVE estimate flows at a given pixel using data from the
neighborhood around that pixel (defined with windowing functions), and
are therefore less susceptible to spurious fluctuations in field
strengths in any individual pixel.

Attempts to ameliorate this shortcoming by smoothing the field
difference $[\Delta B_R]$ rapidly degraded the consistency of the
Poisson solution with the actual magnetic evolution.  This consistency
is checked using Equation (\ref{eqn:induct_dnb}) to calculate the
expected magnetic evolution assuming $\Uvec B_R$ is known ({\it e.g.}
\citealt{Welsch2007}).  Temporal averaging of higher-cadence data,
however, should counteract the effects of spurious, pixel-scale
magnetic fluctuations, but still enable consistency between derived
flows and Equation (\ref{eqn:induct_dnb}).

Unfortunately, statistical agreement between FLCT and Poisson-flow
components is poor, with both linear and rank-order correlations below
$0.1$ for the magnetogram pair in Figure \ref{fig:flct_combo}, both
when weighted by $B_R$ and not.  Proxy Poynting fluxes derived from
tracking and Poisson flow estimates are significantly correlated, but
this could arise solely from spatial correlation of $B_R^2.$ Helicity
fluxes derived from tracking and Poisson flow estimates are also
poorly correlated.  Unlike the proxy Poynting flux, helicity-flux
density depends sensitively upon {\em direction} of flow, since
  $\Uvec$ is dotted with $\avec^P$.  Given that FLCT and Poisson-flow
components are weakly correlated, such disagreement in helicity fluxes
is not surprising.

Given these disagreements between results from each method, which flow
estimate is more credible?  \citet{Welsch2009} found flow estimates
from the FLCT and DAVE methods --- both of which are optical flow
techniques, but derive flow estimates in very different ways
\citep{Schuck2006,Fisher2008} --- were significantly correlated.  If
weighted by $B_R,$ linear correlation coefficients were $> 0.7$ for
each flow component for more than 2000 magnetogram pairs in their
dataset.  In addition, \citet{Welsch2009} found that flow components
derived from either FLCT or DAVE persisted in time, with a
decorrelation time of approximately six hours (see their Figures 9 and
10).  In contrast, Poisson flows for AR 10987 have much lower
frame-to-frame correlations than FLCT flows; they essentially
decorrelate after one 96-minute interval.  Both correlations between
flows determined using the independent FLCT and DAVE methods and the
persistence of FLCT and DAVE flows in time suggest that the FLCT
results found here are more credible than the Poisson flows.

\citet{Welsch2007} tested several velocity-estimation methods using
synthetic magnetogram sequences extracted from MHD simulations, in
which the flows were known. Later, \citet{Welsch2008a} tested
Poisson-flow estimates with the same synthetic data, and found them to
be about as accurate as optical-flow estimates.  Since the tracking
and Poisson flows were both significantly correlated with the known
MHD flows, they were also significantly correlated with each other.
The lack of agreement between FLCT and Poisson flows found here for
96-minute cadence, full-disk MDI magnetogram data suggests the Poisson
method of estimating flows might fail when applied to such data.  The
Poisson technique might, however, fare better when applied to data
with higher cadence and/or higher spatial resolution, such as data
from either MDI in its high-resolution mode or HMI.  Further study of
Poisson flow estimates is planned.

\section{Discussion}
\label{sec:discussion}

By all of the statistical predictors of flare activity based
upon magnetogram structure and evolution that we investigated, AR
10989 should have been significantly less active than both AR 10987
and 10988.  During the intervals that we tracked each AR, however,
flare activity in AR 10989 was comparable to activity in the other
ARs.  Further, \citet{Webb2011} found AR 10989 to be much more CME
productive than the other two ARs.

Why did these statistical predictors fail in the case of the WHI ARs?
Given that we only consider three ARs here, and that these predictors
are purely statistical, we should not be too surprised that there are
exceptions.  Also, flares are rare events, and the time interval of
our observations was limited, so drawing conclusions about activity
levels from the small number of flares observed is problematic.
While we only analyzed the ARs from the WHI for this study, stronger
conclusions regarding relationships (or lack thereof) between ARs'
photospheric magnetic properties and flares could presumably be made
by studying a larger AR sample.
Nonetheless, it would be useful if some physical basis for the
exceptions we found could be identified.  

Unfortunately, the observations available to us are of limited utility
in this regard.  First, a practical concern: as noted above, we have
only LOS magnetogram sequences for the WHI ARs.  Long-duration,
regular-cadence vector magnetogram sequences (such as those available
from HMI, or that might be available from a network of terrestrial,
SOLIS-like instruments) can reveal much more about the structure and
evolution of AR magnetic fields.

Second, another practical concern: we are largely ignorant of the WHI
active regions' magnetic structure and coronal activity over much of
their lives. A significant component of AR 10989's activity occurred
both before and after we tracked it across the central solar disk.
Because the structure and evolution of its photospheric magnetic field
could only be reliably estimated while on the central disk, we can
only speculate about this field away from disk center.
\citet{Welsch2009} showed that AR magnetic fields can evolve
substantially over the course of 24 hours, and \citet{Longcope2005}
and \citet{Schrijver2005} estimated coronal energy storage times on
the order of a day.  If the field of AR 10989 evolved significantly
between its time on the central disk and the pre- and post-tracking
intervals, its photospheric magnetic structure during those more
active invtervals might have been more consistent with statistical
expectations based upon its coronal activity level.
Similarly, we often only have hints about flares and CMEs produced by
ARs beyond the limb.  For instance, STEREO-A data show that on 05
April, AR 10987 at longitude W105 produced a CME just before 16:00UT
with a speed estimated from LASCO data at 962 km s$^{-1}$, associated
with an A-class GOES flare. (See Nitta's excellent compilation of data
for this and other events at
\url{http://www.lmsal.com/nitta/movies/flares_euvi/})
\citet{Burkepile2004} has shown that GOES flux is correlated with CME
kinetic energy, with a linear correlation coefficient of 0.74.
Similarly, the events analyzed by \citet{Qiu2005} show that GOES flux
is correlated with CME speed: we find linear and rank-order
correlation coefficients of 0.79 and 0.54, respectively, in their
sample of 13 events.  These results suggest that the A-class flare
observed by GOES was most likely a partially occulted strong C-class
or weak M-class flare.  To address this ignorance, having more
extensive magnetograph, X-ray, {\em in-situ}, and EUV coverage of the
Sun would be useful.  A satellite in co-rotation with the Sun could
provide comprehensive observations of ARs from birth to death; and a
constellation of satellites might provide full, ``$4\pi$'' coverage.

Third, studying photospheric data alone neglects aspects of coronal
magnetic structure that are relevant to flaring.  For instance, the
large-scale structure of the coronal magnetic field, including perhaps
the proximity of an AR to open magnetic fields, might play a key role
in CME productivity.  \citet{Sterling2001c} argued that open flux near
NOAA AR 8210 enabled repeated, homologous ejections (see their Figure
3).  Similarly, \citet{Liu2007a} reported that CMEs underneath the
streamer belt tended to be substantially slower than CMEs outside the
belt, suggesting that the global magnetic environment around ARs can
play a significant role in the CME process.  Global-scale
potential-field source-surface (PFSS) modeling by \citet{Petrie2011} 
demonstrates that AR 10989 lies almost completely outside the streamer
belt (see their Fig. 21), perhaps easing the ejection of coronal
fields into the heliosphere as CMEs.  Differences in large-scale
coronal magnetic environment might partially explain why flare
activity levels were broadly similar among the WHI ARs, but CME
activity levels were not.  (It is also true that flares and CMEs
appear to be distinct phenomena: very large flares are almost always
associated with CMEs, but that the association is weaker with weaker
flares \citep{Andrews2003}.)  It would be useful to study objective,
quantitative measures of ``proximity'' between open flux and ARs, and
correlations between CME activity and such proximity
measures. \citet{Hudson2010} noted that the flare-to-CME ratio
increased this solar minimum relative to the last, implying the corona
``was relatively easy to disrupt'' during this minimum, also
suggesting global properties of the coronal magnetic field might play
some role in the CME process.

Fourth, properties of photospheric magnetic fields might be
essentially unrelated to properties of the coronal magnetic field that
actually generate flares and CMEs.  Having analyzed a large sample of
vector magnetograms, \citet{Leka2007} suggested ``the state of the
photospheric magnetic field at any given time has limited bearing on
whether that region will be flare productive.''  (Based upon the case
studies here, we might add the evolution of the line-of-sight
photospheric magnetic field over any given time interval also has
limited bearing on whether that region will be flare productive ---
but we have not analyzed enough cases to justify such a conclusion.)
Efforts to measure properties of coronal magnetic fields using IR
(SOLARC: \citealt{Lin2004}; CoMP: \citealt{Tomczyk2009}; ATST:
\citealt{Cargill2009}) and radio (FASR: \citealt{Cargill2009})
instruments could lead to breakthroughs in our understanding of flare
and CME physics.

Finally, leaving aside all observational limitations, flares might be
unpredictable for more fundamental reasons, a possibility also noted
by \citet{Schrijver2009}.

\begin{acks}
 The authors owe thanks to many people: the WHI Team for inviting BTW
 to participate in the Second WHI Workshop; the organizers of this
 Topical Issue of {\it Solar Physics}; the SOHO/MDI and RHESSI teams
 for making their databases available and easy to use; and the
 American taxpayers, for their financial support of this work.  MDI is
 funded through NASA’s Solar and Heliospheric Physics program; SOHO is 
 a project of international cooperation between ESA and NASA.
 This research has made use of NASA's Astrophysics Data System
 Service.  BTW acknowledges support from NSF awards ATM-0752597 and
 AGS-1024862. JMM acknowledges support from NASA grants NAS5-98033,
 NNX08AJ18G and NNX08AI56G.
\end{acks}

\bibliographystyle{/Users/welsch/Library/Latexstuff/SP/spr-mp-sola}


\begin{thebibliography}{52}
\ifx \bisbn   \undefined \def \bisbn  #1{ISBN #1}\fi
\ifx \binits  \undefined \def \binits#1{#1}\fi
\ifx \bauthor  \undefined \def \bauthor#1{#1}\fi
\ifx \batitle  \undefined \def \batitle#1{#1}\fi
\ifx \bjtitle  \undefined \def \bjtitle#1{\textit{#1}}\fi
\ifx \bvolume  \undefined \def \bvolume#1{\textbf{#1}}\fi
\ifx \byear  \undefined \def \byear#1{#1}\fi
\ifx \bissue  \undefined \def \bissue#1{#1}\fi
\ifx \bfpage  \undefined \def \bfpage#1{#1}\fi
\ifx \blpage  \undefined \def \blpage #1{#1}\fi
\ifx \burl  \undefined \def \burl#1{\textsf{#1}}\fi
\ifx \href  \undefined \def \href#1#2{\textsf{#2}}\fi
\ifx \doiurl  \undefined \def
  \doiurl#1{\href{http://dx.doi.org/#1}{\textsf{#1}}}\fi
\ifx \betal  \undefined \def \betal{\textit{et al.}}\fi
\ifx \binstitute  \undefined \def \binstitute#1{#1}\fi
\ifx \bctitle  \undefined \def \bctitle#1{#1}\fi
\ifx \beditor  \undefined \def \beditor#1{#1}\fi
\ifx \bpublisher  \undefined \def \bpublisher#1{#1}\fi
\ifx \bbtitle  \undefined \def \bbtitle#1{\textit{#1}}\fi
\ifx \bedition  \undefined \def \bedition#1{#1}\fi
\ifx \bseriesno  \undefined \def \bseriesno#1{\textbf{#1}}\fi
\ifx \blocation  \undefined \def \blocation#1{#1}\fi
\ifx \bsertitle  \undefined \def \bsertitle#1{\textit{#1}}\fi
\ifx \bsnm \undefined \def \bsnm#1{#1}\fi
\ifx \bsuffix \undefined \def \bsuffix#1{#1}\fi
\ifx \bparticle \undefined \def \bparticle#1{#1}\fi
\ifx \barticle \undefined \def \barticle#1{}\fi
\ifx \botherref \undefined \def \botherref#1{}\fi
\ifx \url \undefined \def \url#1{\textsf{#1}}\fi
\ifx \bchapter \undefined \def \bchapter#1{}\fi
\ifx \bbook \undefined \def \bbook#1{}\fi
\ifx \bcomment \undefined \def \bcomment#1{#1}\fi
\ifx \oauthor \undefined \def \oauthor#1{#1}\fi
\ifx \citeauthoryear \undefined \def \citeauthoryear#1{#1}\fi
\def \endbibitem {}

\bibitem[\protect\citeauthoryear{{Abramenko}}{2005}]{Abramenko2005}
\begin{barticle}
\bauthor{\bsnm{{Abramenko}}, \binits{V.I.}}:
\byear{2005},
\batitle{{Relationship between Magnetic Power Spectrum and Flare Productivity in Solar Active Regions}}.
\bjtitle{Astrophys. J.}
\bvolume{629},
\bfpage{1141}\,--\,\blpage{1149}.
doi:\doiurl{10.1086/431732}.
\end{barticle}
\endbibitem

\bibitem[\protect\citeauthoryear{{Andrews}}{2003}]{Andrews2003}
\begin{barticle}
\bauthor{\bsnm{{Andrews}}, \binits{M.D.}}:
\byear{2003},
\batitle{{A Search for CMEs Associated with Big Flares}}.
\bjtitle{\solphys}
\bvolume{218},
\bfpage{261}\,--\,\blpage{279}.
doi:\doiurl{10.1023/B:SOLA.0000013039.69550.bf}.
\end{barticle}
\endbibitem

\bibitem[\protect\citeauthoryear{{Barnes} and {Leka}}{2008}]{Barnes2008}
\begin{barticle}
\bauthor{\bsnm{{Barnes}}, \binits{G.}}, \bauthor{\bsnm{{Leka}}, \binits{K.D.}}:
\byear{2008},
\batitle{{Evaluating the Performance of Solar Flare Forecasting Methods}}.
\bjtitle{Astrophys. J. Lett.}
\bvolume{688},
\bfpage{L107}\,--\,\blpage{L110}.
doi:\doiurl{10.1086/595550}.
\end{barticle}
\endbibitem

\bibitem[\protect\citeauthoryear{Berger}{1984}]{Berger1984c}
\begin{barticle}
\bauthor{\bsnm{Berger}, \binits{M.A.}}:
\byear{1984},
\batitle{Rigorous new limits on magnetic helicity dissipation in the solar
  corona}.
\bjtitle{Geophys. Astrophys. Fluid Dynamics}
\bvolume{30},
\bfpage{79}\,--\,\blpage{104}.
\end{barticle}
\endbibitem

\bibitem[\protect\citeauthoryear{Berger and Field}{1984}]{Berger1984}
\begin{barticle}
\bauthor{\bsnm{Berger}, \binits{M.A.}}, \bauthor{\bsnm{Field}, \binits{G.B.}}:
\byear{1984},
\batitle{The topological properties of magnetic helicity}.
\bjtitle{J. Fl. Mech.}
\bvolume{147},
\bfpage{133}\,--\,\blpage{148}.
\end{barticle}
\endbibitem

\bibitem[\protect\citeauthoryear{{Burkepile}
  \textit{et~al.}}{2004}]{Burkepile2004}
\begin{barticle}
\bauthor{\bsnm{{Burkepile}}, \binits{J.T.}}, \bauthor{\bsnm{{Hundhausen}},
  \binits{A.J.}}, \bauthor{\bsnm{{Stanger}}, \binits{A.L.}},
  \bauthor{\bsnm{{St.~Cyr}}, \binits{O.C.}}, \bauthor{\bsnm{{Seiden}},
  \binits{J.A.}}:
\byear{2004},
\batitle{{Role of projection effects on solar coronal mass ejection properties:
  1. A study of CMEs associated with limb activity}}.
\bjtitle{J. Geophys. Res. (Space Phys.)}
\bvolume{109}(\bissue{PA18}),
\bfpage{3103}.
doi:\doiurl{10.1029/2003JA010149}.
\end{barticle}
\endbibitem

\bibitem[\protect\citeauthoryear{{Cargill}}{2009}]{Cargill2009}
\begin{barticle}
\bauthor{\bsnm{{Cargill}}, \binits{P.J.}}:
\byear{2009},
\batitle{{Coronal Magnetism: Difficulties and Prospects}}.
\bjtitle{\ssr}
\bvolume{144},
\bfpage{413}\,--\,\blpage{421}.
doi:\doiurl{10.1007/s11214-008-9446-9}.
\end{barticle}
\endbibitem

\bibitem[\protect\citeauthoryear{Chae}{2001}]{Chae2001}
\begin{barticle}
\bauthor{\bsnm{Chae}, \binits{J.}}:
\byear{2001},
\batitle{Observational determination of the rate of magnetic helicity transport
  through the solar surface via the horizontal motion of field line
  footpoints}.
\bjtitle{Astrophys. J. Lett.}
\bvolume{560},
\bfpage{L95}\,--\,\blpage{L98}.
\end{barticle}
\endbibitem

\bibitem[\protect\citeauthoryear{{Christe}, {Krucker}, and
  {Lin}}{2008a}]{Christe2008b}
\begin{barticle}
\bauthor{\bsnm{{Christe}}, \binits{S.}}, \bauthor{\bsnm{{Krucker}},
  \binits{S.}}, \bauthor{\bsnm{{Lin}}, \binits{R.P.}}:
\byear{2008}a,
\batitle{{Hard X-Rays Associated with Type III Radio Bursts}}.
\bjtitle{Astrophys. J. Lett.}
\bvolume{680},
\bfpage{L149}\,--\,\blpage{L152}.
doi:\doiurl{10.1086/589971}.
\end{barticle}
\endbibitem

\bibitem[\protect\citeauthoryear{{Christe}
  \textit{et~al.}}{2008b}]{Christe2008}
\begin{barticle}
\bauthor{\bsnm{{Christe}}, \binits{S.}}, \bauthor{\bsnm{{Hannah}},
  \binits{I.G.}}, \bauthor{\bsnm{{Krucker}}, \binits{S.}},
  \bauthor{\bsnm{{McTiernan}}, \binits{J.}}, \bauthor{\bsnm{{Lin}},
  \binits{R.P.}}:
\byear{2008}b,
\batitle{{RHESSI Microflare Statistics. I. Flare-Finding and Frequency
  Distributions}}.
\bjtitle{\apj}
\bvolume{677},
\bfpage{1385}\,--\,\blpage{1394}.
doi:\doiurl{10.1086/529011}.
\end{barticle}
\endbibitem

\bibitem[\protect\citeauthoryear{{D{\' e}moulin} and
  {Berger}}{2003}]{Demoulin2003}
\begin{barticle}
\bauthor{\bsnm{{D{\' e}moulin}}, \binits{P.}}, \bauthor{\bsnm{{Berger}},
  \binits{M.A.}}:
\byear{2003},
\batitle{{Magnetic Energy and Helicity Fluxes at the Photospheric Level}}.
\bjtitle{Solar Phys.}
\bvolume{215},
\bfpage{203}\,--\,\blpage{215}.
\end{barticle}
\endbibitem

\bibitem[\protect\citeauthoryear{{DeForest}
  \textit{et~al.}}{2007}]{DeForest2007}
\begin{barticle}
\bauthor{\bsnm{{DeForest}}, \binits{C.E.}}, \bauthor{\bsnm{{Hagenaar}},
  \binits{H.J.}}, \bauthor{\bsnm{{Lamb}}, \binits{D.A.}},
  \bauthor{\bsnm{{Parnell}}, \binits{C.E.}}, \bauthor{\bsnm{{Welsch}},
  \binits{B.T.}}:
\byear{2007},
\batitle{{Solar Magnetic Tracking. I. Software Comparison and Recommended
  Practices}}.
\bjtitle{\apj}
\bvolume{666},
\bfpage{576}\,--\,\blpage{587}.
doi:\doiurl{10.1086/518994}.
\end{barticle}
\endbibitem

\bibitem[\protect\citeauthoryear{{Falconer}, {Moore}, and
  {Gary}}{2003}]{Falconer2003}
\begin{barticle}
\bauthor{\bsnm{{Falconer}}, \binits{D.A.}}, \bauthor{\bsnm{{Moore}},
  \binits{R.L.}}, \bauthor{\bsnm{{Gary}}, \binits{G.A.}}:
\byear{2003},
\batitle{{A measure from line-of-sight magnetograms for prediction of coronal
  mass ejections}}.
\bjtitle{J. Geophys. Res. (Space Phys.)}
\bvolume{108}(\bissue{A10}),
\bfpage{11}\,--\,\blpage{1}.
doi:\doiurl{10.1029/2003JA010030}.
\end{barticle}
\endbibitem

\bibitem[\protect\citeauthoryear{{Falconer}, {Moore}, and
  {Gary}}{2006}]{Falconer2006}
\begin{barticle}
\bauthor{\bsnm{{Falconer}}, \binits{D.A.}}, \bauthor{\bsnm{{Moore}},
  \binits{R.L.}}, \bauthor{\bsnm{{Gary}}, \binits{G.A.}}:
\byear{2006},
\batitle{{Magnetic Causes of Solar Coronal Mass Ejections: Dominance of the
  Free Magnetic Energy over the Magnetic Twist Alone}}.
\bjtitle{\apj}
\bvolume{644},
\bfpage{1258}\,--\,\blpage{1272}.
doi:\doiurl{10.1086/503699}.
\end{barticle}
\endbibitem

\bibitem[\protect\citeauthoryear{{Fisher} and {Welsch}}{2008}]{Fisher2008}
\begin{botherref}
\oauthor{\bsnm{{Fisher}}, \binits{G.H.}}, \oauthor{\bsnm{{Welsch}},
  \binits{B.T.}}:
2008,
{FLCT: A Fast, Efficient Method for Performing Local Correlation Tracking}.
In: \textit{Subsurface and Atmospheric Influences on Solar Activity},
{Howe}, R., {Komm}, R.W., {Balasubramaniam}, K.S., {Petrie}, G.J.D. (eds.)
\textit{Astron. Soc. Pacific CS}, San Francisco. 
\textbf{383},
373\,--\,380;\,also\,arXiv:07124289.
\end{botherref}
\endbibitem

\bibitem[\protect\citeauthoryear{{Forbes}}{2000}]{Forbes2000}
\begin{barticle}
\bauthor{\bsnm{{Forbes}}, \binits{T.G.}}:
\byear{2000},
\batitle{{A review on the genesis of coronal mass ejections}}.
\bjtitle{J. Geophys. Res.}
\bvolume{105},
\bfpage{23153}\,--\,\blpage{23166}.
\end{barticle}
\endbibitem

\bibitem[\protect\citeauthoryear{Gosling}{1993}]{Gosling1993}
\begin{barticle}
\bauthor{\bsnm{Gosling}, \binits{J.T.}}:
\byear{1993},
\batitle{The solar flare myth}.
\bjtitle{JGR}
\bvolume{98}(\bissue{A11}),
\bfpage{18937}\,--\,\blpage{18949}.
\end{barticle}
\endbibitem

\bibitem[\protect\citeauthoryear{{Hagenaar}, {Schrijver}, and
  {Title}}{1997}]{Hagenaar1997}
\begin{barticle}
\bauthor{\bsnm{{Hagenaar}}, \binits{H.J.}}, \bauthor{\bsnm{{Schrijver}},
  \binits{C.J.}}, \bauthor{\bsnm{{Title}}, \binits{A.M.}}:
\byear{1997},
\batitle{{The Distribution of Cell Sizes of the Solar Chromospheric Network}}.
\bjtitle{Astrophys. J.}
\bvolume{481},
\bfpage{988}.
doi:\doiurl{10.1086/304066}.
\end{barticle}
\endbibitem

\bibitem[\protect\citeauthoryear{Hagenaar \textit{et~al.}}{1999}]{Hagenaar1999}
\begin{barticle}
\bauthor{\bsnm{Hagenaar}, \binits{H.}}, \bauthor{\bsnm{Schrijver},
  \binits{C.}}, \bauthor{\bsnm{Title}, \binits{A.}}, \bauthor{\bsnm{Shine},
  \binits{R.}}:
\byear{1999},
\batitle{Dispersal of magnetic flux in the quiet solar photosphere}.
\bjtitle{ApJ}
\bvolume{511},
\bfpage{932}\,--\,\blpage{944}.
\end{barticle}
\endbibitem

\bibitem[\protect\citeauthoryear{{Henney} \textit{et~al.}}{2009}]{Henney2009}
\begin{botherref}
\oauthor{\bsnm{{Henney}}, \binits{C.J.}}, \oauthor{\bsnm{{Keller}},
  \binits{C.U.}}, \oauthor{\bsnm{{Harvey}}, \binits{J.W.}},
\oauthor{\bsnm{{Georgoulis}}, \binits{M.K.}},
\oauthor{\bsnm{{Hadder}}, \binits{N.L.}}, \oauthor{\bsnm{{Norton}},
  \binits{A.A.}}, \oauthor{\bsnm{{Raouafi}}, \binits{N.}},
\oauthor{\bsnm{{Toussaint}}, \binits{R.M.}}: 2009, {SOLIS Vector
  Spectromagnetograph: Status and Science}.  In: \textit{SOLAR
  POLARIZATION 5: In Honor of Jan Olof Stenflo}, {S.~V.~Berdyugina,
  K.~N.~Nagendra, R.~Ramelli} (ed.)
\textit{Astron. Soc. Pacific CS}, San Francisco.
\textbf{405},
47.
\end{botherref}
\endbibitem

\bibitem[\protect\citeauthoryear{{Hudson} and {Li}}{2010}]{Hudson2010}
\begin{botherref}
\oauthor{\bsnm{{Hudson}}, \binits{H.S.}}, \oauthor{\bsnm{{Li}}, \binits{Y.}}:
2010,
{Flare and CME Properties and Rates at Sunspot Minimum}.
In: {S.~R.~Cranmer, J.~T.~Hoeksema, J.~L.~Kohl} (ed.)
\textit{SOHO 23: Understanding a Peculiar Solar Minimum},
\textit{Astron. Soc. Pacific CS}, San Francisco. 
\textbf{428},
153.
\end{botherref}
\endbibitem

\bibitem[\protect\citeauthoryear{Kusano \textit{et~al.}}{2002}]{Kusano2002}
\begin{barticle}
\bauthor{\bsnm{Kusano}, \binits{K.}}, \bauthor{\bsnm{Maeshiro}, \binits{T.}},
  \bauthor{\bsnm{Yokoyama}, \binits{T.}}, \bauthor{\bsnm{Sakurai},
  \binits{T.}}:
\byear{2002},
\batitle{Measurement of magnetic helicity injection and free energy loading
  into the solar corona}.
\bjtitle{Astrophys. J.}
\bvolume{577},
\bfpage{501}\,--\,\blpage{512}.
\end{barticle}
\endbibitem

\bibitem[\protect\citeauthoryear{{Leka} and {Barnes}}{2007}]{Leka2007}
\begin{barticle}
\bauthor{\bsnm{{Leka}}, \binits{K.D.}}, \bauthor{\bsnm{{Barnes}}, \binits{G.}}:
\byear{2007},
\batitle{{Photospheric Magnetic Field Properties of Flaring versus Flare-quiet
  Active Regions. IV. A Statistically Significant Sample}}.
\bjtitle{\apj}
\bvolume{656},
\bfpage{1173}\,--\,\blpage{1186}.
\end{barticle}
\endbibitem

\bibitem[\protect\citeauthoryear{{Li} \textit{et~al.}}{2010}]{Li2010}
\begin{barticle}
\bauthor{\bsnm{{Li}}, \binits{Y.}}, \bauthor{\bsnm{{Lynch}}, \binits{B.J.}},
  \bauthor{\bsnm{{Welsch}}, \binits{B.T.}}, \bauthor{\bsnm{{Stenborg}},
  \binits{G.A.}}, \bauthor{\bsnm{{Luhmann}}, \binits{J.G.}},
  \bauthor{\bsnm{{Fisher}}, \binits{G.H.}}, \bauthor{\bsnm{{Liu}},
  \binits{Y.}}, \bauthor{\bsnm{{Nightingale}}, \binits{R.W.}}:
\byear{2010},
\batitle{{Sequential Coronal Mass Ejections from AR8038 in May 1997}}.
\bjtitle{\solphys}
\bvolume{264},
\bfpage{149}\,--\,\blpage{164}.
doi:\doiurl{10.1007/s11207-010-9547-y}.
\end{barticle}
\endbibitem

\bibitem[\protect\citeauthoryear{{Lin}, {Kuhn}, and {Coulter}}{2004}]{Lin2004}
\begin{barticle}
\bauthor{\bsnm{{Lin}}, \binits{H.}}, \bauthor{\bsnm{{Kuhn}}, \binits{J.R.}},
  \bauthor{\bsnm{{Coulter}}, \binits{R.}}:
\byear{2004},
\batitle{{Coronal Magnetic Field Measurements}}.
\bjtitle{Astrophys. J. Lett.}
\bvolume{613},
\bfpage{L177}\,--\,\blpage{L180}.
doi:\doiurl{10.1086/425217}.
\end{barticle}
\endbibitem

\bibitem[\protect\citeauthoryear{{Lin} \textit{et~al.}}{2002}]{Lin2002}
\begin{barticle}
\bauthor{\bsnm{{Lin}}, \binits{R.P.}}, \bauthor{\bsnm{{Dennis}},
  \binits{B.R.}}, \bauthor{\bsnm{{Hurford}}, \binits{G.J.}},
  \bauthor{\bsnm{{Smith}}, \binits{D.M.}}, \bauthor{\bsnm{{Zehnder}},
  \binits{A.}}, \bauthor{\bsnm{{Harvey}}, \binits{P.R.}},
  \bauthor{\bsnm{{Curtis}}, \binits{D.W.}}, \bauthor{\bsnm{{Pankow}},
  \binits{D.}}, \bauthor{\bsnm{{Turin}}, \binits{P.}},
  \bauthor{\bsnm{{Bester}}, \binits{M.}}, \bauthor{\bsnm{{Csillaghy}},
  \binits{A.}}, \bauthor{\bsnm{{Lewis}}, \binits{M.}},
  \bauthor{\bsnm{{Madden}}, \binits{N.}}, \bauthor{\bsnm{{van Beek}},
  \binits{H.F.}}, \bauthor{\bsnm{{Appleby}}, \binits{M.}},
  \bauthor{\bsnm{{Raudorf}}, \binits{T.}}, \bauthor{\bsnm{{McTiernan}},
  \binits{J.}}, \bauthor{\bsnm{{Ramaty}}, \binits{R.}},
  \bauthor{\bsnm{{Schmahl}}, \binits{E.}}, \bauthor{\bsnm{{Schwartz}},
  \binits{R.}}, \bauthor{\bsnm{{Krucker}}, \binits{S.}},
  \bauthor{\bsnm{{Abiad}}, \binits{R.}}, \bauthor{\bsnm{{Quinn}}, \binits{T.}},
  \bauthor{\bsnm{{Berg}}, \binits{P.}}, \bauthor{\bsnm{{Hashii}}, \binits{M.}},
  \bauthor{\bsnm{{Sterling}}, \binits{R.}}, \bauthor{\bsnm{{Jackson}},
  \binits{R.}}, \bauthor{\bsnm{{Pratt}}, \binits{R.}},
  \bauthor{\bsnm{{Campbell}}, \binits{R.D.}}, \bauthor{\bsnm{{Malone}},
  \binits{D.}}, \bauthor{\bsnm{{Landis}}, \binits{D.}},
  \bauthor{\bsnm{{Barrington-Leigh}}, \binits{C.P.}},
  \bauthor{\bsnm{{Slassi-Sennou}}, \binits{S.}}, \bauthor{\bsnm{{Cork}},
  \binits{C.}}, \bauthor{\bsnm{{Clark}}, \binits{D.}}, \bauthor{\bsnm{{Amato}},
  \binits{D.}}, \bauthor{\bsnm{{Orwig}}, \binits{L.}}, \bauthor{\bsnm{{Boyle}},
  \binits{R.}}, \bauthor{\bsnm{{Banks}}, \binits{I.S.}},
  \bauthor{\bsnm{{Shirey}}, \binits{K.}}, \bauthor{\bsnm{{Tolbert}},
  \binits{A.K.}}, \bauthor{\bsnm{{Zarro}}, \binits{D.}},
  \bauthor{\bsnm{{Snow}}, \binits{F.}}, \bauthor{\bsnm{{Thomsen}},
  \binits{K.}}, \bauthor{\bsnm{{Henneck}}, \binits{R.}},
  \bauthor{\bsnm{{McHedlishvili}}, \binits{A.}}, \bauthor{\bsnm{{Ming}},
  \binits{P.}}, \bauthor{\bsnm{{Fivian}}, \binits{M.}},
  \bauthor{\bsnm{{Jordan}}, \binits{J.}}, \bauthor{\bsnm{{Wanner}},
  \binits{R.}}, \bauthor{\bsnm{{Crubb}}, \binits{J.}},
  \bauthor{\bsnm{{Preble}}, \binits{J.}}, \bauthor{\bsnm{{Matranga}},
  \binits{M.}}, \bauthor{\bsnm{{Benz}}, \binits{A.}}, \bauthor{\bsnm{{Hudson}},
  \binits{H.}}, \bauthor{\bsnm{{Canfield}}, \binits{R.C.}},
  \bauthor{\bsnm{{Holman}}, \binits{G.D.}}, \bauthor{\bsnm{{Crannell}},
  \binits{C.}}, \bauthor{\bsnm{{Kosugi}}, \binits{T.}},
  \bauthor{\bsnm{{Emslie}}, \binits{A.G.}}, \bauthor{\bsnm{{Vilmer}},
  \binits{N.}}, \bauthor{\bsnm{{Brown}}, \binits{J.C.}},
  \bauthor{\bsnm{{Johns-Krull}}, \binits{C.}}, \bauthor{\bsnm{{Aschwanden}},
  \binits{M.}}, \bauthor{\bsnm{{Metcalf}}, \binits{T.}},
  \bauthor{\bsnm{{Conway}}, \binits{A.}}:
\byear{2002},
\batitle{{The Reuven Ramaty High-Energy Solar Spectroscopic Imager (RHESSI)}}.
\bjtitle{\solphys}
\bvolume{210},
\bfpage{3}\,--\,\blpage{32}.
doi:\doiurl{10.1023/A:1022428818870}.
\end{barticle}
\endbibitem

\bibitem[\protect\citeauthoryear{{Linker} \textit{et~al.}}{2003}]{Linker2003}
\begin{barticle}
\bauthor{\bsnm{{Linker}}, \binits{J.A.}}, \bauthor{\bsnm{{Miki{\'c}}},
  \binits{Z.}}, \bauthor{\bsnm{{Lionello}}, \binits{R.}},
  \bauthor{\bsnm{{Riley}}, \binits{P.}}, \bauthor{\bsnm{{Amari}}, \binits{T.}},
  \bauthor{\bsnm{{Odstrcil}}, \binits{D.}}:
\byear{2003},
\batitle{{Flux cancellation and coronal mass ejections}}.
\bjtitle{Physics of Plasmas}
\bvolume{10},
\bfpage{1971}\,--\,\blpage{1978}.
doi:\doiurl{10.1063/1.1563668}.
\end{barticle}
\endbibitem

\bibitem[\protect\citeauthoryear{{Liu}}{2007}]{Liu2007a}
\begin{barticle}
\bauthor{\bsnm{{Liu}}, \binits{Y.}}:
\byear{2007},
\batitle{{Halo Coronal Mass Ejections and Configuration of the Ambient Magnetic
  Fields}}.
\bjtitle{\apjl}
\bvolume{654},
\bfpage{L171}\,--\,\blpage{L174}.
doi:\doiurl{10.1086/511385}.
\end{barticle}
\endbibitem

\bibitem[\protect\citeauthoryear{{Liu}, {Norton}, and
  {Scherrer}}{2007}]{Liu2007b}
\begin{barticle}
\bauthor{\bsnm{{Liu}}, \binits{Y.}}, \bauthor{\bsnm{{Norton}}, \binits{A.A.}},
  \bauthor{\bsnm{{Scherrer}}, \binits{P.H.}}:
\byear{2007},
\batitle{{A Note on Saturation Seen in the MDI/SOHO Magnetograms}}.
\bjtitle{\solphys}
\bvolume{241},
\bfpage{185}\,--\,\blpage{193}.
doi:\doiurl{10.1007/s11207-007-0296-5}.
\end{barticle}
\endbibitem

\bibitem[\protect\citeauthoryear{{Livi}, {Wang}, and {Martin}}{1985}]{Livi1985}
\begin{barticle}
\bauthor{\bsnm{{Livi}}, \binits{S.H.B.}}, \bauthor{\bsnm{{Wang}}, \binits{J.}},
  \bauthor{\bsnm{{Martin}}, \binits{S.F.}}:
\byear{1985},
\batitle{{The cancellation of magnetic flux. I - On the quiet sun}}.
\bjtitle{Austral. J. Phys.}
\bvolume{38},
\bfpage{855}\,--\,\blpage{873}.
\end{barticle}
\endbibitem

\bibitem[\protect\citeauthoryear{Longcope \textit{et~al.}}{2005}]{Longcope2005}
\begin{barticle}
\bauthor{\bsnm{Longcope}, \binits{D.W.}}, \bauthor{\bsnm{McKenzie},
  \binits{D.}}, \bauthor{\bsnm{Cirtain}, \binits{J.}}, \bauthor{\bsnm{Scott},
  \binits{J.}}:
\byear{2005},
\batitle{{Observations of separator reconnection to an emerging active
  region}}.
\bjtitle{\apj}
\bvolume{630},
\bfpage{596}\,--\,\blpage{614}.
\end{barticle}
\endbibitem

\bibitem[\protect\citeauthoryear{{Low}}{2001}]{Low2001}
\begin{barticle}
\bauthor{\bsnm{{Low}}, \binits{B.C.}}:
\byear{2001},
\batitle{{Coronal mass ejections, magnetic flux ropes, and solar magnetism}}.
\bjtitle{\jgr}
\bvolume{106},
\bfpage{25141}\,--\,\blpage{25164}.
doi:\doiurl{10.1029/2000JA004015}.
\end{barticle}
\endbibitem

\bibitem[\protect\citeauthoryear{{Martin}}{1998}]{Martin1998}
\begin{barticle}
\bauthor{\bsnm{{Martin}}, \binits{S.F.}}:
\byear{1998},
\batitle{{Conditions for the Formation and Maintenance of Filaments - (Invited
  Review)}}.
\bjtitle{Solar Phys.}
\bvolume{182},
\bfpage{107}\,--\,\blpage{137}.
\end{barticle}
\endbibitem

\bibitem[\protect\citeauthoryear{Pariat, Demoulin, and
  Berger\,}{2005}]{Pariat2005}
\begin{barticle}
\bauthor{\bsnm{Pariat}, \binits{E.}}, \bauthor{\bsnm{Demoulin}, \binits{P.}},
  \bauthor{\bsnm{Berger}, \binits{M.A.}}:
\byear{2005},
\batitle{{Photospheric flux density of magnetic helicity}}.
\bjtitle{\aap}
\bvolume{439},
\bfpage{1191}\,--\,\blpage{1203}.
doi:\doiurl{10.1051/0004-6361:20052663}.
\end{barticle}
\endbibitem

\bibitem[\protect\citeauthoryear{{Parnell} \textit{et~al.}}{2009}]{Parnell2009}
\begin{barticle}
\bauthor{\bsnm{{Parnell}}, \binits{C.E.}}, \bauthor{\bsnm{{Deforest}},
  \binits{C.E.}}, \bauthor{\bsnm{{Hagenaar}}, \binits{H.J.}},
  \bauthor{\bsnm{{Johnston}}, \binits{B.A.}}, \bauthor{\bsnm{{Lamb}},
  \binits{D.A.}}, \bauthor{\bsnm{{Welsch}}, \binits{B.T.}}:
\byear{2009},
\batitle{{A Power-law Distribution of Solar Magnetic Fields Over More Than Five
  Decades in Flux}}.
\bjtitle{\apj}
\bvolume{698},
\bfpage{75}\,--\,\blpage{82}.
\end{barticle}
\endbibitem

\bibitem[\protect\citeauthoryear{{Petri,Canou, and Amari}\,}{2011}]{Petrie2011}
\begin{barticle}
\bauthor{\bsnm{Petrie}, \binits{G.J.D.}}, \bauthor{\bsnm{Canou}, \binits{A.}},
  \bauthor{\bsnm{Amari}, \binits{T.}}:
\byear{2011},
\batitle{{Nonlinear Force-Free and Potential-Field Models of Active-Region and 
Global Coronal Fields during the Whole Heliosphere Interval}}.
\bjtitle{Solar Phys.}
\bvolume{Online First},
\bfpage{}\,--\,\blpage{}.
doi:\doiurl{10.1007/s11207-010-9687-0}.
\end{barticle}
\endbibitem


\bibitem[\protect\citeauthoryear{Pevtsov, Canfield, and
  McClymont}{1997}]{Pevtsov1997}
\begin{barticle}
\bauthor{\bsnm{Pevtsov}, \binits{A.A.}}, \bauthor{\bsnm{Canfield},
  \binits{R.C.}}, \bauthor{\bsnm{McClymont}, \binits{A.N.}}:
\byear{1997},
\batitle{On the subphotospheric origin of coronal electric currents}.
\bjtitle{Astrophys. J.}
\bvolume{481},
\bfpage{973}\,--\,\blpage{977}.
\end{barticle}
\endbibitem

\bibitem[\protect\citeauthoryear{{Qiu} and {Yurchyshyn}}{2005}]{Qiu2005}
\begin{barticle}
\bauthor{\bsnm{{Qiu}}, \binits{J.}}, \bauthor{\bsnm{{Yurchyshyn}},
  \binits{V.B.}}:
\byear{2005},
\batitle{{Magnetic Reconnection Flux and Coronal Mass Ejection Velocity}}.
\bjtitle{Astrophys. J. Lett.}
\bvolume{634},
\bfpage{L121}\,--\,\blpage{L124}.
doi:\doiurl{10.1086/498716}.
\end{barticle}
\endbibitem

\bibitem[\protect\citeauthoryear{Scherrer \textit{et~al.}}{1995}]{Scherrer1995}
\begin{barticle}
\bauthor{\bsnm{Scherrer}, \binits{P.}}, \bauthor{\bsnm{Bogart}, \binits{R.S.}},
  \bauthor{\bsnm{Bush}, \binits{R.I.}}, \bauthor{\bsnm{Hoeksema},
  \binits{J.T.}}, \bauthor{\bsnm{Kosovichev}, \binits{A.}},
  \bauthor{\bsnm{Schou}, \binits{J.}}, \bauthor{\bsnm{Rosenberg}, \binits{W.}},
  \bauthor{\bsnm{Springer}, \binits{L.}}, \bauthor{\bsnm{Tarbell},
  \binits{T.}}, \bauthor{\bsnm{Title}, \binits{A.}}, \bauthor{\bsnm{Wolfson},
  \binits{C.}}, \bauthor{\bsnm{Zayer}, \binits{I.}}, \bauthor{\bsnm{{The MDI
  Engineering Team}}}:
\byear{1995},
\batitle{The solar oscillations investigation - michelson doppler imager}.
\bjtitle{Solar~Phys.}
\bvolume{162},
\bfpage{129}\,--\,\blpage{188}.
\end{barticle}
\endbibitem

\bibitem[\protect\citeauthoryear{{Schrijver}}{2007}]{Schrijver2007}
\begin{barticle}
\bauthor{\bsnm{{Schrijver}}, \binits{C.J.}}:
\byear{2007},
\batitle{{A Characteristic Magnetic Field Pattern Associated with All Major
  Solar Flares and Its Use in Flare Forecasting}}.
\bjtitle{Astrophys. J. Lett.}
\bvolume{655},
\bfpage{L117}\,--\,\blpage{L120}.
doi:\doiurl{10.1086/511857}.
\end{barticle}
\endbibitem

\bibitem[\protect\citeauthoryear{{Schrijver}}{2009}]{Schrijver2009}
\begin{barticle}
\bauthor{\bsnm{{Schrijver}}, \binits{C.J.}}:
\byear{2009},
\batitle{{Driving major solar flares and eruptions: A review}}.
\bjtitle{Adv. Space Res.}
\bvolume{43},
\bfpage{739}\,--\,\blpage{755}.
doi:\doiurl{10.1016/j.asr.2008.11.004}.
\end{barticle}
\endbibitem

\bibitem[\protect\citeauthoryear{{Schrijver}
  \textit{et~al.}}{2005}]{Schrijver2005}
\begin{barticle}
\bauthor{\bsnm{{Schrijver}}, \binits{C.J.}}, \bauthor{\bsnm{{DeRosa}},
  \binits{M.L.}}, \bauthor{\bsnm{{Title}}, \binits{A.M.}},
  \bauthor{\bsnm{{Metcalf}}, \binits{T.R.}}:
\byear{2005},
\batitle{{The Nonpotentiality of Active-Region Coronae and the Dynamics of the
  Photospheric Magnetic Field}}.
\bjtitle{\apj}
\bvolume{628},
\bfpage{501}\,--\,\blpage{513}.
\end{barticle}
\endbibitem

\bibitem[\protect\citeauthoryear{{Schuck}}{2006}]{Schuck2006}
\begin{barticle}
\bauthor{\bsnm{{Schuck}}, \binits{P.W.}}:
\byear{2006},
\batitle{{Tracking Magnetic Footpoints with the Magnetic Induction Equation}}.
\bjtitle{\apj}
\bvolume{646},
\bfpage{1358}\,--\,\blpage{1391}.
doi:\doiurl{10.1086/505015}.
\end{barticle}
\endbibitem

\bibitem[\protect\citeauthoryear{{Schuck}}{2008}]{Schuck2008}
\begin{barticle}
\bauthor{\bsnm{{Schuck}}, \binits{P.W.}}:
\byear{2008},
\batitle{{Tracking Vector Magnetograms with the Magnetic Induction Equation}}.
\bjtitle{\apj}
\bvolume{683},
\bfpage{1134}\,--\,\blpage{1152}.
doi:\doiurl{10.1086/589434}.
\end{barticle}
\endbibitem

\bibitem[\protect\citeauthoryear{{Sterling} and {Moore}}{2001}]{Sterling2001c}
\begin{barticle}
\bauthor{\bsnm{{Sterling}}, \binits{A.C.}}, \bauthor{\bsnm{{Moore}},
  \binits{R.L.}}:
\byear{2001},
\batitle{{Internal and external reconnection in a series of homologous solar
  flares}}.
\bjtitle{\jgr}
\bvolume{106},
\bfpage{25227}\,--\,\blpage{25238}.
doi:\doiurl{10.1029/2000JA004001}.
\end{barticle}
\endbibitem

\bibitem[\protect\citeauthoryear{{Tomczyk} and {McIntosh}}{2009}]{Tomczyk2009}
\begin{barticle}
\bauthor{\bsnm{{Tomczyk}}, \binits{S.}}, \bauthor{\bsnm{{McIntosh}},
  \binits{S.W.}}:
\byear{2009},
\batitle{{Time-Distance Seismology of the Solar Corona with CoMP}}.
\bjtitle{\apj}
\bvolume{697},
\bfpage{1384}\,--\,\blpage{1391}.
doi:\doiurl{10.1088/0004-637X/697/2/1384}.
\end{barticle}
\endbibitem

\bibitem[\protect\citeauthoryear{{Wang} \textit{et~al.}}{2009}]{Wang2009}
\begin{barticle}
\bauthor{\bsnm{{Wang}}, \binits{D.}}, \bauthor{\bsnm{{Zhang}}, \binits{M.}},
  \bauthor{\bsnm{{Li}}, \binits{H.}}, \bauthor{\bsnm{{Zhang}}, \binits{H.Q.}}:
\byear{2009},
\batitle{{A Cross-Comparison of Cotemporal Magnetograms Obtained with MDI/SOHO
  and SP/ Hinode}}.
\bjtitle{\solphys}
\bvolume{260},
\bfpage{233}\,--\,\blpage{244}.
doi:\doiurl{10.1007/s11207-009-9441-7}.
\end{barticle}
\endbibitem

\bibitem[\protect\citeauthoryear{{Webb} \textit{et~al.}}{2011}]{Webb2011}
\begin{barticle}
\bauthor{\bsnm{{Webb}}, \binits{D.F.}}, 
  \bauthor{\bsnm{{Cremades}}, \binits{H.}},
  \bauthor{\bsnm{{Sterling}}, \binits{A.C}}, 
  \bauthor{\bsnm{{Mandrini}}, \binits{C.H.}}, 
  \bauthor{\bsnm{{Dasso}}, \binits{S.}}, 
  \bauthor{\bsnm{{Gibson}}, \binits{S.E.}}, 
  \bauthor{\bsnm{{Haber}}, \binits{D.A.}}, 
  \bauthor{\bsnm{{Komm}}, \binits{R.W.}}, 
  \bauthor{\bsnm{{Petrie}}, \binits{G.J.D.}}, 
  \bauthor{\bsnm{{McIntosh}}, \binits{P.S.}}, 
  \bauthor{\bsnm{{Welsch}}, \binits{B.T.}}, 
  \bauthor{\bsnm{{Simon}}, \binits{S.P.}}:
\byear{20011},
\batitle{{The Global Context of Solar Activity During the Whole Heliosphere
Interval Campaign}}.
\bjtitle{\solphys}
\bvolume{(in preparation)},
\bfpage{}\,--\,\blpage{}.
\end{barticle}
\endbibitem

\bibitem[\protect\citeauthoryear{Welsch}{2006}]{Welsch2006}
\begin{barticle}
\bauthor{\bsnm{Welsch}, \binits{B.T.}}:
\byear{2006},
\batitle{Magnetic flux cancellation and coronal magnetic energy}.
\bjtitle{Astrophys. J.}
\bvolume{638},
\bfpage{1101}.
\end{barticle}
\endbibitem

\bibitem[\protect\citeauthoryear{{Welsch} and {Fisher}}{2008}]{Welsch2008a}
\begin{botherref}
\oauthor{\bsnm{{Welsch}}, \binits{B.T.}}, \oauthor{\bsnm{{Fisher}},
  \binits{G.H.}}:
2008,
{Surface Flows From Magnetograms}.
In: \textit{Subsurface and Atmospheric Influences on Solar Activity},
{Howe}, R., {Komm}, R.W., {Balasubramaniam}, K.S., {Petrie}, G.J.D. (eds.)
\textit{Astron. Soc. Pacific CS}, San Francisco. 
\textbf{383},
19\,--\,30;\,also\,arXiv:07100546.
\end{botherref}
\endbibitem

\bibitem[\protect\citeauthoryear{{Welsch} and {Li}}{2008}]{Welsch2008b}
\begin{botherref}
\oauthor{\bsnm{{Welsch}}, \binits{B.T.}}, \oauthor{\bsnm{{Li}}, \binits{Y.}}:
2008,
{On the Origin of Strong-Field Polarity Inversion Lines}.
In: \textit{Subsurface and Atmospheric Influences on Solar Activity},
{Howe}, R., {Komm}, R.W., {Balasubramaniam}, K.S., {Petrie}, G.J.D. (eds.)
\textit{Astron. Soc. Pacific CS}, San Francisco. 
\textbf{383},
429\,--\,437;\,also\,arXiv:07100562.
\end{botherref}
\endbibitem

\bibitem[\protect\citeauthoryear{Welsch and Longcope}{2003}]{Welsch2003}
\begin{botherref}
\oauthor{\bsnm{Welsch}, \binits{B.T.}}, \oauthor{\bsnm{Longcope},
  \binits{D.W.}}:
2003,
Magnetic helicity injection by horizontal flows in the quiet sun: I. mutual
  helicity flux.
\textit{Astrophys. J.}
\textbf{588}.
\end{botherref}
\endbibitem

\bibitem[\protect\citeauthoryear{Welsch \textit{et~al.}}{2004}]{Welsch2004}
\begin{barticle}
\bauthor{\bsnm{Welsch}, \binits{B.T.}}, \bauthor{\bsnm{Fisher}, \binits{G.H.}},
  \bauthor{\bsnm{Abbett}, \binits{W.P.}}, \bauthor{\bsnm{R\'egnier},
  \binits{S.}}:
\byear{2004},
\batitle{{ILCT}: Recovering photospheric velocities from magnetograms by
  combining the induction equation with local correlation tracking}.
\bjtitle{Astrophys. J.}
\bvolume{610},
\bfpage{1148}.
\end{barticle}
\endbibitem

\bibitem[\protect\citeauthoryear{Welsch \textit{et~al.}}{2007}]{Welsch2007}
\begin{barticle}
\bauthor{\bsnm{Welsch}, \binits{B.T.}}, \bauthor{\bsnm{Abbett}, \binits{W.P.}},
  \bauthor{\bsnm{DeRosa}, \binits{M.L.}}, \bauthor{\bsnm{Fisher},
  \binits{G.H.}}, \bauthor{\bsnm{Georgoulis}, \binits{K.}
  \bsuffix{M.~K.~Kusano}}, \bauthor{\bsnm{Longcope}, \binits{D.W.}},
  \bauthor{\bsnm{Ravindra}, \binits{B.}}, \bauthor{\bsnm{Schuck},
  \binits{P.W.}}:
\byear{2007},
\batitle{Tests and comparisons of velocity inversion techniques}.
\bjtitle{Astrophys. J.}
\bvolume{670},
\bfpage{1434}\,--\,\blpage{1452}.
\end{barticle}
\endbibitem

\bibitem[\protect\citeauthoryear{{Welsch} \textit{et~al.}}{2009}]{Welsch2009}
\begin{barticle}
\bauthor{\bsnm{{Welsch}}, \binits{B.T.}}, \bauthor{\bsnm{{Li}}, \binits{Y.}},
  \bauthor{\bsnm{{Schuck}}, \binits{P.W.}}, \bauthor{\bsnm{{Fisher}},
  \binits{G.H.}}:
\byear{2009},
\batitle{{What is the Relationship Between Photospheric Flow Fields and Solar
  Flares?}}
\bjtitle{\apj}
\bvolume{705},
\bfpage{821}\,--\,\blpage{843}.
doi:\doiurl{10.1088/0004-637X/705/1/821}.
\end{barticle}
\endbibitem

\end{thebibliography}

\end{article}
 
\end{document}